\documentclass[12pt]{article}
\usepackage{amscd,amssymb,amsmath,latexsym,enumerate}
\usepackage[mathscr]{euscript}
\usepackage{mathrsfs}
\usepackage{epsfig}
\usepackage{fancybox}
\usepackage{verbatim}
\usepackage{tikz}
\usepackage{tikz-cd}
\usepackage{todonotes}
\usepackage{multicol}
\usepackage{graphicx}

\usepackage{color}

\usepackage{xcolor}
\definecolor{MyBlue}{cmyk}{1,0.13,0,0.63}
\definecolor{MyGreen}{cmyk}{0.91,0,0.88,0.52}
\newcommand{\mylinkcolor}{MyBlue}
\newcommand{\mycitecolor}{MyGreen}
\newcommand{\myurlcolor}{black}

\usepackage{hyperref}
\hypersetup{%
  bookmarksnumbered=true,bookmarksopen=false,%
  plainpages=false,
  linktocpage=true,%
  colorlinks=true,breaklinks=true,%
  linkcolor=\mylinkcolor,citecolor=\mycitecolor,urlcolor=\myurlcolor,%
  pdfpagelayout=OneColumn,%
  pageanchor=true,%
}

\title{Spectral localization for semimetals and Callias operators 
}

\author{Hermann Schulz-Baldes and Tom Stoiber
\\
\\
{\small  Friedrich-Alexander-Universit\"at Erlangen-N\"urnberg}
\\
{\small Department Mathematik, Cauerstr. 11, D-91058 Erlangen, Germany}
\\
{\small Email: schuba@mi.uni-erlangen.de, tom.stoiber@fau.de}
}

\vspace{.5cm}

\date{ }

\newtheorem{theorem}{Theorem}[section]
\newtheorem{proposition}[theorem]{Proposition}
\newtheorem{lemma}[theorem]{Lemma}

\newtheorem{definition}[theorem]{Definition}

\numberwithin{equation}{section}

\newcommand{\CM}{{\mathbb C}}

\newcommand{\RM}{{\mathbb R}}
\newcommand{\SM}{{\mathbb S}}
\newcommand{\TM}{{\mathbb T}}
\newcommand{\ZM}{{\mathbb Z}}

\newcommand{\FM}{{\mathbb F}}

\newcommand{\Ee}{{\cal E}}

\newcommand{\Bb}{{\cal B}}

\newcommand{\Ff}{{\cal F}}

\newcommand{\Vv}{{\cal V}}
\newcommand{\Ss}{{\cal S}}
\newcommand{\Oo}{{\cal O}}

\newcommand{\Mm}{{\cal M}}

\newcommand{\Hh}{{\cal H}}

\newcommand{\Zz}{{\cal Z}}

\newcommand{\one}{{\bf 1}}

\newcommand{\Tr}{\mbox{\rm Tr}}
\newcommand{\ev}{{\mbox{\tiny\rm ev}}}
\newcommand{\od}{{\mbox{\tiny\rm od}}}

\newcommand{\SF}{{\rm Sf}}

\newcommand{\Ch}{{\rm Ch}} 
\newcommand{\Ind}{{\rm Ind}} 
\newcommand{\Ker}{{\rm Ker}} 
\newcommand{\Ran}{{\rm Ran}} 
 
\newcommand{\sgn}{{\rm sgn}} 
\newcommand{\Sig}{{\rm Sig}} 
\newcommand{\diag}{{\rm diag}}

\newcommand{\sa}{{\mbox{\rm\tiny sa}}}
\newcommand{\ssa}{{\mbox{\rm\tiny ssa}}}
\newcommand{\mult}{q}

\begin{document}

\maketitle

\begin{abstract}
A semiclassical argument is used to show that the low-lying spectrum of a selfadjoint operator, the so-called spectral localizer, determines the number of Dirac or Weyl points of an ideal semimetal.  Apart from the IMS localization procedure, an explicit computation for the local toy models given by a Dirac or Weyl point is the key element of proof. The argument has numerous similarities to Witten's reasoning leading to the strong Morse inequalities. The same techniques allow to prove a spectral localization for Callias operators associated to potentials with isolated gap-closing points. 




\end{abstract}



\section{Introduction}

The main motivation of this paper is to provide the analytical justification for an earlier numerical study on the Weyl point count in semimetals via a spectral localizer \cite{SSt1}.  More precisely, let $H$ be a one-particle periodic tight-binding Hamiltonian describing an ideal semimetal in odd dimension $d$ with $I$ Weyl points, see Definition~\ref{def-IdealSemimet} below for details. If $\widehat{D}=\sum_{j=1}^dX_j\otimes\gamma_j$ is the dual Dirac operator expressed in terms of the position operator and $\sigma_1$, $\sigma_2$ and $\sigma_3$ are the Pauli matrices, then the
the spectral localizer with tuning parameter $\kappa>0$ is defined as $L_\kappa=\kappa \widehat{D}\otimes \sigma_1-H\otimes \sigma_3$. It is proved that its low-lying spectrum consists of exactly $I$ eigenvalues that are well-separated from the rest of the spectrum: 
$$
\Tr\big(
\chi_{[-c\kappa^{\frac{1}{2}},c\kappa^{\frac{1}{2}}]}(L_\kappa)\big)
\;=\;
I
\;=\;
\Tr\big(
\chi_{[-C\kappa^{\frac{2}{3}},C\kappa^{\frac{2}{3}}]}(L_\kappa)\big)
\;,
$$
for two constants $c$ and $C$. Here $\chi_K$ is the indicator function on a subset $K\subset\RM$. The proof combines a spectral localization estimate in the semiclassical parameter $\kappa$ with an explicit computation of the contributions stemming from the spectral localizer of a Weyl Hamiltonian. For even dimension $d$ there is an analogous result on the Dirac point count.

\vspace{.2cm}

The spectral localizer is the supersymmetric version of the Callias-type operator $\kappa \widehat{D}+\imath H$ on the compact manifold $\TM^d$.  The spectral localization techniques directly transpose to Callias operators on the non-compact manifold $\RM^d$ {having a finite zero set}, even if the Callias potential $H$ takes values in the selfadjoint Fredholm operators. {This leads a spectral localization for the Callias index theorem which can also be interpreted as an index theorem for a non-commutative version of spectral flow (in the sense of \cite{Wahl2008,KL,vD,Kub,SSt})}, thus generalizing results of Robbin and Salamon \cite{RoS} and Abbondandolo and Majer \cite{AM}. {These results on Callias operators also illustrate the index-theoretic aspects of the result on semimetals.} Sections~\ref{sec-Overview} and \ref{sec-Overview2} contain the full stories on semimetals and Callias operators respectively, mathematical details and proofs are then provided in the remainder of the paper.

\section{Results on semimetals} 
\label{sec-Overview}

\subsection{Semimetals as topological transition points} 
\label{sec-SemimetalTransition}

The prototypical and probably most studied topological quantum system is the Harper Hamiltonian $H_\theta$ on the two-dimensional lattice $\ZM^2$ depending on the magnetic flux $\theta$ through each lattice cell. Its spectrum is well-known to consist of absolutely continuous bands for rational $\theta$, while it is fractal for irrational $\theta$. To energies outside of the spectrum or more precisely to the spectral projection below such an energy, one can associate a Chern number \cite{TKN2,Bel,BES}. On the other hand, a given energy can become a point of band touching at which the Chern number is not well-defined. Such an energy can rather be seen as a transition point between two different topological states. This typically happens at rational values of $\theta$ for which one then has a transfer of topological charge from the lower to the upper band \cite{Bel95}. If the Fermi surface consists of only such points, the density of states vanishes there and one also speaks of a semimetal, namely in these models there is no gap as in an insulator, but merely a pseudo-gap at the Fermi level where the density of states vanishes. 

\vspace{.2cm}

Such a transmission of topological charge at a semimetallic transition point also appears at so-called Dirac points in higher even dimensions  \cite[Section 2.2.4]{PSbook} and is often also called a topological phase transition in the physics literature. A similar scenario is also well-known to happen for one-parameter families of periodic topological insulators in odd dimensions \cite[Section 2.3.3]{PSbook}. Then the Hamiltonian $H$ of such a family is supposed to have a chiral symmetry $\Gamma_0 H\Gamma_0=-H$ with $\Gamma_0^*=\Gamma_0$ and $\Gamma_0^2=\one$, and the transitions between topological states with different (higher) winding numbers are through so-called Weyl points. The most simple example is the one-dimensional SSH-model in which changes of the mass term lead to such transitions. Even though this is not the main focus of this work, we include Section~\ref{sec-TopoMass} on such transitions points (mathematical details on it are then given in Section~\ref{sec-ChernIntegrals}).

\vspace{.2cm}

Apart from transition points between topological insulators, semimetal behavior is also generic for periodic Schr\"odinger operators with supplementary symmetries. For example, it is generic for two-dimensional chiral symmetric models. The graphene Hamiltonian is a model in this class and genericity means that changes of the parameters of the model conserving the chiral symmetry do not destroy the Dirac points ({\it e.g.} \cite{Dro}). For the discussion of higher dimensional models, let us recall a theorem of von Neumann and Wigner \cite{NW} stating that double degeneracy of eigenvalues of a family of self-adjoint matrices indexed by real parameters is an event of real codimension $3$, and higher degeneracies have even higher codimension. In dimension $d=3$, this implies that double degeneracies of the spectrum are generic and these lead to conical intersections of the energy bands called Weyl points, but the corresponding energies of these points are generically not pinned to the same value, thus the band-structure can still be metallic. {This work, however, focusses on models for which the Weyl points are pinned at the same energy and there are no other bands at the Fermi level. Such models will be called ideal semimetals  (this terminology is also used in \cite{Van}). Hence the Fermi surface of an ideal semimetal consists of a finite number of points and its density of states vanishes at the Fermi level (producing what is called a pseudogap). While this is not generic for arbitrary three-dimensional periodic models, it is an open condition in the set Hamiltonians having inversion symmetry. Otherwise stated, it is a stable property to have an inversion symmetric three-dimensional Hamiltonian with two Weyl points with opposite topological charge (this is often called a magnetic Weyl semimetal).} If on top of the inversion symmetry the three-dimensional Hamiltonian has a time-reversal symmetry, then the conical points are four-fold degenerate and are rather called Dirac points, since the linearization of the band-structure corresponds to a three-dimensional Dirac Hamiltonian (which is equivalent to a degenerate pair of two Weyl points with opposite chirality ({\it e.g.} \cite[Section 5.4]{Van}). Below in Section~\ref{sec-TopoMass} it will be shown that, after a doubling procedure, it is possible to add suitable mass terms to any symmetry stabilized semimetal to open a gap and produce a topological transition point.

\vspace{.2cm}

Let us now provide a more formal definition of {an ideal semimetal.} Roughly stated, it is a periodic system for which the Fermi level $E_F=0$ intersects the band-structure at only finite many singular points, around whom the linearization takes the form of a Dirac or Weyl Hamiltonian. We restrict to periodic systems without magnetic fields as periodic rational magnetic fluxes lead to the same set-up (with an increased dimension of the matrix fiber).

\begin{definition}
\label{def-IdealSemimet}
Let $H=P(S_1,\ldots,S_d)$ be a periodic tight-binding Hamiltonian on $\ell^2(\ZM^d,\CM^N)$, namely a matrix-valued polynomial in the lattice shift operators $S_1,\ldots,S_d$. It hence has a Bloch representation over the { (flat) Brillouin torus $\TM^d=\RM^d/\ZM^d\cong [0,1)^d$} which after Bloch-Floquet {\rm (}here simply Fourier{\rm )} transform is given by
\begin{equation}
\label{eq-HFourier}
\Ff H\Ff^*
\;=\;
\int^\oplus_{\TM^d} dk\,{H}_k
\;,
\end{equation}
where $k\in\TM^d\mapsto H_k\in\CM^{N\times N}$ is a real analytic function of selfadjoint matrices. Recall that the Fermi surface at Fermi energy $0$ is the set $\Zz(H)=\{k\in\TM^d\,:\,\dim(\Ker(H_k))\geq 1\}$. Then $H$ is called an ideal Weyl semimetal for odd $d$ respectively an ideal Dirac semimetal for even $d$ provided the following conditions hold:
\begin{enumerate}
\item[{\rm (i)}]  The Fermi surface $\Zz(H)$ is finite, {\it i.e.} $\Zz(H)=\{k^*_1,\ldots,k^*_I\}$.
\label{eq-SingPoints}
\item[{\rm (ii)}] 
For each $k^*\in\Zz(H)$ there exists a ball ${B}_\delta(k^*)$ of some size $\delta>0$ around $k^*$ on which there exists a unitary basis change $k\in{B}_\delta(k^*)\mapsto W_k\in \mbox{\rm U}(N)$ such that
\begin{equation}
\label{eq-WeylPoint}
W_{k} {H}_kW_{k}^*
\;=\;
\begin{pmatrix}
{H}_{k}^0 & 0 
\\
0 & {H}^Q_{k}+ {H}^R_{k}
\end{pmatrix}
\;,
\end{equation}
with an invertible matrix ${H}_{k}^0$, a remainder ${H}^R_{k}$ satisfying $\|{H}^R_{k}\|\leq C|k-k^*|^2$, and linear term ${H}^Q_{k}$ given by a direct sum of $\mult^*$ summands of the form 
\begin{equation}
\label{eq-WeylIntro}
{H}^{W/D}_{k}
\;=\;
{\sum_{j=1}^d \langle  k-k^*| B e_j\rangle\, \Gamma_j}
\;,
\end{equation}
%
{with a real invertible matrix $B\in\RM^{d\times d}$ and where $e_1,\ldots,e_d\in\RM^d$ are the standard basis vectors, the scalar product is in $\RM^d$} and finally $\Gamma_1,\ldots,\Gamma_d$ is an irreducible representation of the complex Clifford algebra $\CM_d$ with $d$ generators, which is hence of dimension $d'=2^{\lfloor\frac{d}{2}\rfloor}$, {and which is supposed to be left-handed for $d$ odd in the sense that $\Gamma_1\cdots\Gamma_d = \imath^{\frac{d-1}{2}}\one$}. If $d$ is odd, then $k^*$ is called a Weyl point and the upper index $W$ is used; on the other hand for $d$ even, it is called a Dirac point and the index $D$ is used. If $k^*=k^*_i$ carries an index $i=1,\ldots,I$, so will the associated  objects ${H}_{i,k}^0$,  ${H}^{Q}_{i,k}$ and {$B_{i}$} etc.. In particular, $\mult^*_i$ denotes the multiplicity of the singular point $k^*_i$.
\end{enumerate}
\end{definition}

As $({H}^{(*)}_{k})^2={\sum_{j=1}^d \langle  k-k^*| B e_j\rangle^2}$, condition (ii) implies that for each $k^*\in\Zz(H)$ exists $c > 0 $ such that $\|H_k\| \geq c |k-k^*|$ in a neighborhood of $k^*$, notably there is a linear vanishing of the density of states at $E_F$. For $d=3$, this corresponds to so-called Type I Weyl semimetals, and Type II Weyl semimetal in the sense of \cite{Van} are not further considered here. Let us also note that a twice degenerate Weyl point with opposite chiralities (which are defined further down) is often also called a Dirac point in the physics literature {(and also exists in odd dimension $d$)}.

\vspace{.2cm}

The set-up described in Definition~\ref{def-IdealSemimet} is stable {(under perturbations)} in dimension $d=2$ and $d=3$ { when the degeneracy is minimal $d'=2$}, provided that the symmetries described above hold ({\it i.e.} chiral symmetry in $d=2$ and inversion symmetry in $d=3$). {In particular, the Clifford algebra structures appear naturally at a band-touching point with linear dispersion, since the crossing of two bands is locally described by a compression $k \mapsto {H}^Q_{k}+ {H}^R_{k}\in \CM^{2\times 2}$ and the Pauli matrices are a basis of the traceless self-adjoint $2\times 2$ matrices.} However, for $d\geq 4$ the singular points have higher degeneracies $d'>2$ which can be perturbed to singular hypersurfaces of codimension $3$ with two-fold degeneracies (in accordance with the theorem of von Neumann and Wigner \cite{NW}). A typical (and stable) example is the Hamiltonian $H=\Gamma_1(-\Delta_{d-2}-\mu)+\Gamma_2\,\frac{1}{\imath}\partial_{d-1}+\Gamma_3\,\frac{1}{\imath}\partial_d$ on $L^2(\RM^d,\CM^2)$ where $\Gamma_1$, $\Gamma_2$ and $\Gamma_3$ are simply the Pauli matrices. Then the singular surface for this example is the determined by $(\sum_{j=1}^{d-2}k_j^2-\mu)^2+k_{d-1}^2+k_d^2=0$ and thus given by a $(d-3)$-sphere. Even though non-generic, {we can only deal with ideal semimetals in the above restrictive sense  for $d>3$. } 
Let us provide two concrete examples which were studied numerically in \cite{SSt1}.

\vspace{.2cm}

\noindent {\bf Example} of a $d=2$ chiral Dirac semimetal: As already stated, the (chiral) graphene Hamiltonian provides a standard two-dimensional model with two Dirac points. Alternatively, one can use a stacked SSH model of the form
$$
H
\;=\;
\begin{pmatrix}
0 & S_1-\delta(S_2+S_2^*)-\mu
\\
S^*_1-\delta(S_2+S_2^*)-\mu & 0
\end{pmatrix}
\;,
$$
which acts on $\ell^2(\ZM^2,\CM^2)$ and is expressed in terms of the shifts $S_1$ and $S_2$ in the two spacial direction. It contains the SSH model in the $1$-direction and hence the periodic two-dimensional model $H$ has Dirac points if $2\delta\cos(k_2)+\mu=\pm 1$. For $4\delta<2$ and $\mu\in(-1-2\delta,1+2\delta)$, there are two Dirac points, while for $4\delta>2$ and $\mu$ sufficiently small, there are $4$ Dirac points. 
\hfill $\diamond$

\vspace{.2cm}

\noindent {\bf Example} of an ideal  Weyl semimetal in dimension $d=3$: 
A minimal model for an ideal Weyl semimetal with merely two Weyl points is given in \cite{ArmitageEtAl} (see eq. (12) therein, but we choose parameters and energy shifts a bit differently):
$$
H
\,=\,
\begin{pmatrix}
\sum_{i=1,2,3}(S_i+S_i^*)-\mu & \delta\big((S_1-S_1^*)+\imath (S_2-S_2^*)\big)
\\
\delta\big((S_1^*-S_1)+\imath (S_2-S_2^*)\big) & -\sum_{i=1,2,3}(S_i+S_i^*)+\mu
\end{pmatrix}
\,.
$$
This contains a stacked $p+ip$ model  in the $(1,2)$-plane
$$
H_{p+ip}
\;=\;
\begin{pmatrix}
\sum_{i=1,2}(S_i+S_i^*)-\hat{\mu} & \delta\big((S_1-S_1^*)+\imath (S_2-S_2^*)\big)
\\
\delta\big((S_1^*-S_1)+\imath (S_2-S_2^*)\big) & -\sum_{i=1,2}(S_i+S_i^*)+\hat{\mu}
\end{pmatrix}
\;,
$$
with a kinetic coupling in the $3$-direction as well as an energy shift $2\eta\sin(k_3)$. For $\delta>0$, the two-dimensional model $H_{p+ip}$ has a band touching at $E=0$ for $\hat{\mu}\in\{-4,0,4\}$  with one Dirac point for $|\hat{\mu}|=4$ and two for $\hat{\mu}=0$. Hence in the periodic three-dimensional model this leads to Weyl points for $\mu-2\cos(k_3)\in\{-4,0,4\}$. For $\mu\in
(-6,-2)\cup(2,6)$, there are two values $k_z$ and two corresponding Weyl points, while for $\mu\in(-2,2)$ there are four Weyl points. It is possible to add {an inversion symmetry breaking term} $\imath\eta\sum_{i=1}^3(S_i-S_i^*)\otimes\one$ which shifts the Weyl points energetically, so that one does not have an ideal semimetal with a pseudo-gap any more.
\hfill $\diamond$

\subsection{Strong topological invariants via the spectral localizer} 

The main new result of the paper is not on the analysis of topological transition points, but rather on a refined (actually semiclassical) analysis of the spectral localizer \cite{LS1,LS2} at such transition points, or - in other terms - on the use of the spectral localizer for semimetals. We recall from \cite{LS1,LS2} and review below that the spectral localizer is a selfadjoint operator constructed from a Hamiltonian and a dual Dirac operator whose spectral asymmetry provides a marker for strong topological invariants. There are also further generalizations and variants of the spectral localizer which have been used to compute $\ZM_2$-invariants \cite{DoS2}, spin Chern numbers and other integer invariants associated to approximate symmetries or conservation laws \cite{DoS} and weak invariants \cite{SSt0}. It has also been suggested to use the spectral localizer to determine spectral gaps \cite{Lor} and as markers in topological metals \cite{CL}.

\vspace{.2cm}

Consider a short-range and periodic tight-binding Hamiltonian $H$ on $\ell^2(\ZM^d,\CM^N)$.  If $H$ is gapped at the Fermi level $E_F=0$ and $d$ is even, then the flat band Hamiltonian $Q=\sgn(H)$ (or equivalently the Fermi projection $P=\chi(H<0)=\frac{1}{2}(\one-Q)$) has a well-defined and integer strong even Chern number $\Ch_d(Q)=\Ch_d(Q,\TM^d)\in\ZM$. For even $d>0$, it is defined by
\begin{align}
\label{eq-EvenChern}
\Ch_{d}(Q,\Mm^{d}) 
\;=\; 
-\,\frac{1}{2}\,
\Big(
\frac{\imath}{8 \pi} \Big)^{\frac{d}{2}}
\frac{1}{ \frac{d}{2}!} 
\int_{\Mm^{d}} 
\Tr\big(Q(dQ)^{\wedge d}\big)
\; ,
\qquad
d\;\mbox{even}\,,
\end{align}
where $\Mm^d$ is a $d$-dimensional smooth {oriented and closed} manifold or simplex and $k\in\Mm^d\mapsto Q_k$ a smooth function of symmetries, namely $Q_k^*=Q_k$ and $(Q_k)^2=\one$. For odd dimension $d$, one supposes that $H$ has, moreover, a chiral  (or sublattice) symmetry $\Gamma_0 H\Gamma_0 =-H$ w.r.t. to a symmetry operator $\Gamma_0$. Then $H$ is off-diagonal in the eigenbasis of $\Gamma_0$:
\begin{equation}
\label{eq-ChiralHam}
H
\;=\;
\begin{pmatrix}
0 & A^* \\ A & 0
\end{pmatrix}
\;,
\qquad
\Gamma_0
\;=\;
\begin{pmatrix} 
\one & 0 \\ 0 & -\one
\end{pmatrix}
\;,
\end{equation}
and $A$ is invertible if $H$ is gapped. {Note that $A:\Ker(\Gamma_0-\one) \to \Ker(\Gamma_0+\one)$ maps the positive spectral subspace of $\Gamma_0$ to the negative spectral subspace of $\Gamma_0$, so that there is no sign ambiguity in the following.} Moreover, also $\Gamma_0 Q\Gamma_0=-Q$. Then $Q$ (or $A$) has a well-defined odd Chern number $\Ch_d(Q)=\Ch_{d}(Q,\TM^{d})\in\ZM$ which is defined by
\begin{align}
\label{eq-OddChern}
\Ch_{d}(Q,\Mm^{d}) 
\;=\; 
-\,
\Big(
\frac{1}{4\pi\imath} \Big)^{\frac{d+1}{2}}
\frac{1}{d!!} 
\int_{\Mm^{d}} 
\Tr\big(\Gamma_0 Q(dQ)^{\wedge d}\big)
\; ,
\qquad
d\;\mbox{odd}\,,
\end{align}
where still $\Mm^d$ is a $d$-dimensional smooth manifold or simplex and $k\in\Mm^d\mapsto Q_k$ a differentiable function of symmetries satisfying $\Gamma_0 Q_k\Gamma_0=-Q_k$. Moreover, $d!!=d(d-2)\cdots 3\cdot 1$. For both $d$ even and odd, $\Ch_d(Q)$ is called the strong topological invariant. The reader is  referred to \cite{PSbook} for non-commutative versions of these invariants as well as their properties.

\vspace{.2cm}

The strong Chern numbers can be computed as the half-signature of a so-called spectral localizer \cite{LS1,LS2}. These latter two papers distinguish the odd and even spectral localizer for odd and even dimension respectively. Let us here propose a spectral localizer that unites both of them. The idea is to start out for both even and odd dimension from the $d$-dimensional dual Dirac operator
\begin{equation}
\label{eq-DiracDef}
\widehat{D}
\;=\;
\sum_{j=1}^d\gamma_j X_j
\;,
\end{equation}
with $X_1,\ldots,X_d$ being the components of the {selfadjoint} position operator and $\gamma_1,\ldots,\gamma_d$ being a {left-handed} irreducible representation of the $d$-dimensional Clifford algebra $\CM_d$, {where the left-handedness here means that as in Definition~\ref{def-IdealSemimet} one has  $\gamma_1\cdots\gamma_d = \imath^{\frac{d-1}{2}}\one$ for odd $d$ (note that this holds for the standard Pauli matrices).} It is a self-adjoint operator on the Hilbert space $\ell^2(\ZM^d,\CM^{d'})$ where $\CM^{d'}$ is the representation space of $\CM_d$. After discrete Fourier transform, it becomes the Dirac operator $D$ on the torus:
\begin{equation}
\label{eq-DiracDef2}
D
\;=\;
\Ff\widehat{D}\Ff^*
\;=\;
-\,\imath\sum_{j=1}^d\gamma_j \partial_j
\;.
\end{equation}
If now $H$ is a short-range tight-binding Hamiltonian acting on $\ell^2(\ZM^d,\CM^N)$ so that $\|[H,X_j]\|\leq C<\infty$ for $j=1,\ldots,d$, the associated spectral localizer is by definition
\begin{equation}
\label{eq-GenSpecLoc}
L_\kappa
\;=\;
\begin{pmatrix}
-H & \kappa\,\widehat{D} \\ \kappa\,\widehat{D} & H
\end{pmatrix}
\;,
\end{equation}
where $\kappa>0$ is a tuning parameter  and we have tacitly written $H$ for {$H\otimes \one_{d'}$} and $\widehat{D}$ for {$\widehat{D}\otimes \one_N$}. If $\sigma_1$, $\sigma_2$ and $\sigma_3$ denote the standard Pauli matrices in the grading of \eqref{eq-GenSpecLoc}, then one has
$$
L_\kappa
\;=\;
\kappa \widehat{D}\otimes\sigma_1\,-\,H\otimes\sigma_3
\;.
$$
The spectral localizer is a selfadjoint operator on {$\ell^2(\ZM^d,\CM^{d'}\otimes \CM^N \otimes \CM^2)$} with compact resolvent. Its low-lying spectrum roughly coincides with the spectrum of (sufficiently large) finite volume restrictions of the spectral localizer, namely its restriction with Dirichlet boundary condition to the range of $\chi(\widehat{D}^*\widehat{D}\leq \rho^2)=\chi(\widehat{D}\widehat{D}^*\leq \rho^2)$. These restrictions are denoted by $L_{\kappa,\rho}$ and are finite-dimensional selfadjoint matrices. To apply the main results of \cite{LS1,LS2} the tuning parameter $\kappa$ and system size $\rho$ are required to satisfy the bounds
\begin{equation}
\label{eq-LocalizerCond}
\kappa
\;<\;
\frac{g^3}{12\|H\|^3\,\|[\widehat{D},H]\|}
\;,
\qquad
\rho 
\;>\;
\frac{2g}{\kappa}
\;,
\end{equation}
where $g=\|H^{-1}\|^{-1}$ is the spectral gap of $H$. These bounds imply that $L_{\kappa,\rho}$ is invertible with gap $\|L_{\kappa,\rho}^{-1}\|^{-1}>\frac{g}{2}$ (this is proved exactly as in \cite{LS1,LS2}).  Let us note that due to its special form \eqref{eq-GenSpecLoc} the spectral localizer has a chiral symmetry (or supersymmetry)
\begin{equation}
\label{eq-ChiralGenSL}
\sigma_2\,L_\kappa\,
\sigma_2
\;=\;
-\,L_\kappa
\;,
\end{equation}
implying that its spectrum is symmetric. In particular, it has no spectral asymmetry and the signature of $L_{\kappa,\rho}$ vanishes. In the grading of $\sigma_2$ which is obtained essentially via the Cayley transform, $L_\kappa$ is then off-diagonal, namely
\begin{equation}
\label{eq-SpecLocCrep}
d^*\,L_\kappa\,d
\;=\;
\begin{pmatrix}
0 & \kappa\,\widehat{D}-\imath\,H
\\
\kappa\,\widehat{D}+\imath\,H & 0
\end{pmatrix}
\;,
\qquad
d\;=\;
\frac{1}{\sqrt{2}}
\begin{pmatrix}
1 & \imath \\ \imath & 1
\end{pmatrix}
\;,
\end{equation}
because $d^*\sigma_1 d=\sigma_1$, $d^*\sigma_2 d=\sigma_3$ and $d^*\sigma_3 d=-\sigma_2$. The lower off-diagonal operator is the Callias operator further discussed in Section~\ref{sec-Overview2}.

\vspace{.2cm}

Now, if either $\widehat{D}$ or $H$ or both have another  symmetry, $L_\kappa$ may be block diagonal and the resulting blocks are then the odd and even spectral localizers of \cite{LS1} and \cite{LS2} respectively and they may have a spectral asymmetry which is precisely linked to the strong topological invariant by the results of \cite{LS1,LS2}. The symmetries involved are actually chiral (anticommuting with $\widehat{D}$ or $H$, hence often also called anti-symmetries), and they are complex in the sense that they do not involve any real structure. Hence the spectral localizer is actually associated to pairings of complex $K$-theory with complex $K$-homology (no algebra is specified here, but one may simply take the commutative algebra generated by $H$ to merge with the terminology of non-commutative geometry \cite{Con,GVF}). More precisely, for odd dimension $d$ the Dirac operator specifies an odd $K^1$-homology class, while for even $d$ it is rather an even $K^0$-homology class, namely for even $d$ it is well-known that for any irreducible representation  $\gamma_1,\ldots,\gamma_d$ of $\CM_d$ there exists {a chirality operator $\gamma_{0}=(-\imath)^{d/2}\gamma_1\cdots\gamma_d$} anti-commuting with all generators. {It is chosen such that $(\gamma_1,\ldots,\gamma_d,\gamma_0)$ is a left-handed irreducible representation of $\CM_{d+1}$.} Let the representation be such that $\gamma_{0}=\binom{1 \;\;0}{0\;-1}$ is given by the third Pauli matrix. Then in its grading $\widehat{D}$ is thus off-diagonal with off-diagonal entries $\widehat{D}_0$ and $\widehat{D}_0^*$: 
\begin{equation}
\label{eq-DiracEvenRep}
\widehat{D}
\;=\;
\begin{pmatrix}
0 & \widehat{D}_0^* \\ \widehat{D}_0 & 0
\end{pmatrix}
\;.
\end{equation}
On the other hand, the Hamiltonian $H$ specifies an even $K_0$-class (once again: of an algebra not specified here, see \cite{PSbook} for details) via its Fermi projection $P=\chi(H<0)$ if $H$ has no further symmetry, while if $H$ has a chiral (or sublattice) symmetry $\Gamma_0 H\Gamma_0 =-H$ it fixes an odd $K_1$-class $A$ which is given by \eqref{eq-ChiralHam}. Let us now go through the four cases separately.

\vspace{.2cm}

\noindent {\bf Case of both $H$ and $\widehat{D}$ without symmetry ($d$ odd; pairing of $K_0$ with $K^1$):}  The spectral localizer given by \eqref{eq-GenSpecLoc} has only the symmetry \eqref{eq-ChiralGenSL}. While this spectral localizer has no spectral asymmetry, it will be used to detect the number of Weyl points in odd-dimensional physical systems, see Section~\ref{sec-StatementMain}.

\vspace{.2cm}

\noindent {\bf Case of $H$ without and $\widehat{D}$ with chiral symmetry ($d$ even; pairing of $K_0$ with $K^0$):}  
As $[\gamma_{0},H]=0$, one consequently also has $\gamma_{0} L_\kappa\gamma_{0}=-L_\kappa$. Combined with \eqref{eq-ChiralGenSL} the spectral localizer $L_\kappa$ is a $4\times 4$ matrix splitting into a direct sum:
\begin{equation}
\label{eq-PairingEvenEven}
L_\kappa
\;=\;
\begin{pmatrix}
-H & \kappa\,\widehat{D}_0^* \\ \kappa\,\widehat{D}_0 & H
\end{pmatrix}
\oplus
\begin{pmatrix}
-H & \kappa\,\widehat{D}_0 \\ \kappa\,\widehat{D}_0^* & H
\end{pmatrix}
\;.
\end{equation}
The first summand is the even spectral localizer from \cite{LS2}:
\begin{equation}
\label{eq-EvenSpecLoc}
L_\kappa^\ev
\;=\;
\begin{pmatrix}
-H & \kappa\,\widehat{D}_0^* \\ \kappa\,\widehat{D}_0 & H
\end{pmatrix}
\;.
\end{equation}
It can have a spectral asymmetry and its half-signature is equal to the index pairing \cite{LS2}. More precisely, if $\kappa$ and $\rho$ satisfy \eqref{eq-LocalizerCond}, then
\begin{equation}
\label{eq-evenSLForm}
\Ch_d(Q)
\;=\;
\frac{1}{2}\,\Sig(L^\ev_{\kappa,\rho})
\;,
\end{equation}
where $\Sig(L^\ev_{\kappa,\rho})$ denotes the signature of the selfadjoint and invertible matrix $L^\ev_{\kappa,\rho}$. Provided that the $\eta$-invariant of $L^\ev_\kappa$ exists, one can replace $\Sig(L_{\kappa,\rho}^\ev)$ by $\eta(L_{\kappa}^\ev)$ in \eqref{eq-evenSLForm}, see \cite{LS1}. Of course, the signature and $\eta$-invariant of the second summand in \eqref{eq-PairingEvenEven} are the same up to a sign change. Let us also stress that \eqref{eq-evenSLForm} as well as \eqref{eq-oddSLForm} below also hold and are proved for covariant media in the sense of \cite{BES,PSbook}, but this is not relevant for the discussion here. Finally, let us anticipate that in an even-dimensional ideal semimetal $L^\ev_\kappa$ can be used to count the number of Dirac points of $H$ in even dimension, see Section~\ref{sec-StatementMain}.

\vspace{.2cm}

\noindent {\bf Case of $H$ with and $\widehat{D}$ without chiral symmetry ($d$ odd; pairing of $K_1$ with $K^1$):} Now the Hamiltonian is of the form \eqref{eq-ChiralHam}. As $[\Gamma_0 ,\widehat{D}]=0$, the spectral localizer is again a direct sum, most conveniently written in the form
\begin{equation}
\label{eq-PassageOddEven0}
e\,L_{\kappa}\,e^*
\;=\;
\begin{pmatrix}
\kappa \widehat{D} & A^* \\ A & -\kappa \widehat{D}
\end{pmatrix}
\oplus
\begin{pmatrix}
\kappa \widehat{D} & A \\ A^* & -\kappa \widehat{D}
\end{pmatrix}
\;,
\qquad
e\;=\;
\frac{1}{\sqrt{2}}
\begin{pmatrix} 1 & 1 \\ -1 & 1 \end{pmatrix}
\;,
\end{equation}
where $e$ is a unitary basis change in the matrix degrees of freedom of \eqref{eq-GenSpecLoc} which actually satisfies $e\sigma_1e^*=\sigma_3$, $e\sigma_2 e^*=\sigma_2$ and $e\sigma_3 e^*=-\sigma_1$. The first summand is the odd spectral localizer used in \cite{LS1}:
\begin{equation}
\label{eq-oddSL}
L^\od_\kappa
\;=\;
\begin{pmatrix}
\kappa\,\widehat{D} & A^* \\  A & -\kappa\,\widehat{D}
\end{pmatrix}
\;.
\end{equation}
The main result of \cite{LS1} states that, provided $\kappa$ and $\rho$ satisfy \eqref{eq-LocalizerCond}, one has for $Q=H|H|^{-1}$
\begin{equation}
\label{eq-oddSLForm}
\Ch_d(Q)
\;=\;
-\,\frac{1}{2}\,\Sig(L^\od_{\kappa,\rho})
\;.
\end{equation}

\vspace{.2cm}

\noindent {\bf Case of both $H$ and $\widehat{D}$ with chiral symmetry ($d$ even; pairing of $K_1$ with $K^0$):}  In this case, one has the spectral localizer is as in \eqref{eq-PairingEvenEven}, but moreover the even spectral localizer $L^\ev_\kappa$ conjugated by $\Gamma\otimes\sigma_1$ is equal to the second summand in \eqref{eq-PairingEvenEven}. This implies that neither summand in \eqref{eq-PairingEvenEven} can have a spectral asymmetry.  On the other hand, replacing $H$ by \eqref{eq-ChiralHam} shows $L^\ev_\kappa$ is a $4\times 4$ matrix with $8$ vanishing entries. Similar as in \eqref{eq-SpecLocCrep}, it can be brought in an off-diagonal supersymmetric form:
\begin{equation}
\label{eq-SpecLocSSeven}
a\,L^\ev_\kappa\,a^*
\;=\;
\begin{pmatrix}
0 & 0 & -A^* & \kappa \widehat{D}_0^*
\\
0 & 0 &   \kappa \widehat{D}_0 & A 
\\
-A & \kappa \widehat{D}_0^* &  0 & 0 
\\
\kappa  \widehat{D}_0 & A^* & 0 & 0
\end{pmatrix}
\;,
\qquad
a
\;=\;
\begin{pmatrix}
1 & 0 & 0 & 0
\\
0 & 0 & 0 & 1 
\\
0 & 1 & 0 & 0 
\\
0 & 0 & 1 & 0 
\end{pmatrix}
\;.
\end{equation}
The lower off-diagonal entry is the even Callias operator that will be used in Section~\ref{sec-IntroEven}.

\subsection{Motivation for main result} 

In order to provide some motivation and intuition for the result in the next section, let us consider a transition point in a system of even dimension $d$. Hence consider a one-parameter family $m\mapsto H(m)$ of periodic tight-binding Hamiltonians on $\ell^2(\ZM^d,\CM^N)$ having a topological transition point at $m=0$. In particular, $H(m)$ is gapped at the Fermi level $E_F=0$ for $|m|\in(0,m_0)$, but has no gap for $m=0$. Then the flat band Hamiltonian $Q(m)=\sgn(H(m))$ (or equivalently the Fermi projection) has a well-defined and integer strong Chern number $\Ch_d(Q(m))$ for small $m\not=0$ which according to \eqref{eq-evenSLForm} can be computed from the even spectral localizer
\begin{equation}
\label{eq-EvenSpecLocbis}
\Ch_d(Q(m))
\;=\;
\frac{1}{2}\;\Sig(L^\ev_{\kappa,\rho}(m))
\;,
\qquad
L^\ev_\kappa(m)
\;=\;
\begin{pmatrix}
-H(m) & \kappa\,\widehat{D}_0^* \\ \kappa\,\widehat{D}_0 & H(m)
\end{pmatrix}
\;.
\end{equation}
It follows that the transfer of topological charge $\Ch_d(Q(m))-\Ch_d(Q(-m))$ results from a change $\frac{1}{2}\,\Sig(L^\ev_{\kappa,\rho}(m))-\frac{1}{2}\,\Sig(L^\ev_{\kappa,\rho}(-m))$ of half-signature or a change of the (half) $\eta$-invariants, notably a change of the spectral asymmetry of the even spectral localizer for any $\rho$ sufficiently large. On the other hand, $m\mapsto L^\ev_{\kappa}(m)$ is a path of selfadjoint operators with compact resolvent and therefore \cite{APS,Get,LS1} the change of spectral asymmetry is given by the spectral flow of this family, namely one concludes that
\begin{equation}
\label{eq-ChernDiffSF}
\Ch_d(Q(m))\,-\,\Ch_d(Q(-m))
\;=\;
\frac{1}{2}\;
\SF\big(
m'\in[-m,m]\mapsto L^\ev_{\kappa}(m')
\big)
\;.
\end{equation}
This identity holds for all $\kappa$ sufficiently small so that  the first condition in \eqref{eq-LocalizerCond} holds. These prior results therefore imply that there are eigenvalues of $m'\mapsto L^\ev_{\kappa}(m')$ crossing $0$, but is not possible to infer when this happens. For symmetry reasons, one may expect that the eigenvalue crossings leading to the spectral flow in \eqref{eq-ChernDiffSF} happen precisely at $m'=0$ and, if this is true, that $L^\ev_{\kappa}(0)$, the spectral localizer of the transition point,  actually has a kernel of even dimension. Let us stress that this latter fact cannot be {deduced} from the earlier result \cite{LS2} because for any fixed $\kappa$ the gap is not allowed to close due to the first condition in \eqref{eq-LocalizerCond}.

\subsection{Statement of main result on ideal semimetals} 
\label{sec-StatementMain}

The main result of this paper confirms that the expectation on the low-lying spectrum of the spectral localizer of an ideal semimetal as stated at the end of the last section holds, at least up to error terms.

\begin{theorem}
\label{theo-SemiclassicsIntro}
Let $H$ {be} an ideal semimetal with $I$ singular points at the Fermi level $0$. For $d$ odd, the approximate kernel dimension of the associated  spectral localizer $L_\kappa$ given by \eqref{eq-GenSpecLoc} is equal to the number of singular points counted with their multiplicity. More precisely, there are constants $c$ and $C$ such that the spectrum of $L_\kappa$ in the interval $[-c\kappa^{\frac{2}{3}},c\kappa^{\frac{2}{3}}]$ consists of $\sum_{i=1}^I\mult^*_i$ eigenvalues and there is no further spectrum in $[-C\kappa^{\frac{1}{2}},C\kappa^{\frac{1}{2}}]$. In particular, 
$$
\sum_{i=1}^I\mult^*_i
\;=\;
\Tr\big(\chi(|L_\kappa|\leq c\kappa^{\frac{2}{3}})\big)
\;.
$$
If $d$ is even and $H$ an ideal Dirac semimetal, the same results hold for the even spectral localizer $L^\ev_\kappa$ given by \eqref{eq-EvenSpecLoc}.
\end{theorem}

Using the techniques of \cite{LS1,LS2}, one can moreover show that the low-lying spectrum of $L_\kappa$ and its finite-volume restrictions $L_{\kappa,\rho}$ to boxes of size $\rho>0$ coincide up to errors terms that vanish as $\rho\to\infty$. This allows to access the low-lying spectrum of $L_\kappa$ numerically. {This allows to determine the semimetallic nature of a possibly complicated tight-binding model obtained from quantum chemistry. Indeed, if the spectrum of the spectral localizer has an approximate kernel of the form stated in Theorem~\ref{theo-SemiclassicsIntro}, one can even predict the number of Weyl or Dirac points .} 
Such a numerical illustration of Theorem~\ref{theo-SemiclassicsIntro} for models in $d=2$ and $d=3$ was provided in \cite{SSt1}. Numerics also indicate that the low-lying eigenvalues are actually in a window much smaller than $\kappa^{\frac{2}{3}}$. Based on a semiclassical picture, the next section explains what is likely the true $\kappa$-dependence on the eigenvalues. The size $\kappa^{\frac{1}{2}}$ of the larger window gives an optimal scale.

\subsection{Semiclassical perspective on the spectral localizer} 
\label{sec-RoughSemiclassicalPicture}

The proof of Theorem~\ref{theo-SemiclassicsIntro} is essentially the same for even and odd dimension. Hence let us here focus on $d$ odd. The statement of  Theorem~\ref{theo-SemiclassicsIntro} is about counting the low-lying eigenvalues of $L_\kappa$, or equivalently the lowest eigenvalues of its non-negative square $(L_\kappa)^2$ given by
$$
(L_\kappa)^2
\;=\;
\begin{pmatrix}
H^2 + \kappa^2\,\widehat{D}^2 & \kappa [\widehat{D},H] \\ \kappa [H,\widehat{D}] & H^2 + \kappa^2\,\widehat{D}^2
\end{pmatrix}
\;.
$$
After discrete Fourier transform this operator becomes a Schr\"odinger operator on $L^2(\TM^d,\CM^{2N{d'}})$:
\begin{equation}
\label{eq-GenSpecLocSquare}
\Ff(L_\kappa)^2\Ff^*
\;=\;
-\,\kappa^2\,\Delta\;+\;
\begin{pmatrix}
(H_k)^2 & \kappa ({\gamma\cdot\partial_kH_k}) \\ \kappa  ({\gamma\cdot\partial_kH_k}) & (H_k)^2
\end{pmatrix}
\;,
\end{equation}
where $\Delta=-D^2$ is the Laplace operator on the torus and the second summand is a matrix-valued  potential given in terms of the coefficient matrixes in \eqref{eq-HFourier}. Note that $\kappa$ plays the role of Planck's constant $\hbar$, but that other than in standard situations also the potential depends on it. Further note that Theorem~\ref{theo-SemiclassicsIntro} is actually about the semiclassical regime of small values of $\kappa$. Let us give an intuitive description on how to obtain the low-lying spectrum of $(L_\kappa)^2$, supposing for sake of simplicity that each singular point of $k\in\TM^d\mapsto H_k$ is simple in the sense that it has only one Weyl point, {\it i.e.} $\mult^*_i=1$. The matrix potential is given by $k\in\TM^d\mapsto (H_k)^2\otimes \one+\Oo(\kappa)$ and hence has, due to \eqref{eq-WeylPoint}, two harmonic wells precisely at the singular points $k^*_1,\ldots,k^*_I$ up to higher order corrections in $\kappa$, one for the upper and one for the lower component in \eqref{eq-GenSpecLocSquare}. Each of these wells leads to a ground state of order $\frac{1}{2}\kappa$ and excited states at a distance of order $\kappa$, but actually the off-diagonal terms {of} order $\kappa$ split the two-fold fundamental and pushes one eigenvalue down to $0$ and the other up to order $\kappa$. Summing up, for each singular point $k^*_i$ there is precisely one vector in the kernel of $(L_\kappa)^2$ and the remainder of the spectrum lies above $\kappa$. Theorem~\ref{theo-SemiclassicsIntro} is a weakened form of this claim.

\vspace{.2cm}

The first element of the formal proof is the semiclassical localization procedure via the IMS formula \cite{Sim,CFKS}, actually a version of a matrix-valued extension put forward by Shubin \cite{Shu}. This leads to a local toy model for each singular point given by the spectral localizer of a Weyl Hamiltonian or a Dirac Hamiltonian, pending on whether the dimension is odd or even. Its spectrum can be computed explicitly by algebraic means, see Sections~\ref{sec-Weyl} and \ref{sec-SpecDocDiracPoints} respectively. These toy models play the role of the harmonic oscillator Hamiltonian or the saddle point Hamiltonians in semiclassical spectral analysis, see, in particular, the work of Helffer and Sj\"ostrand on the Witten Laplacian \cite{HS0} and on the Harper equation (see \cite{HS} and references therein). However, to our best knowledge the elementary analysis in Sections~\ref{sec-Weyl} and \ref{sec-SpecDocDiracPoints} is novel.

\vspace{.2cm}

Now let us now comment on what we expect to be the true behavior of one of the eigenvalues $\nu_i(\kappa)$ on $\kappa$. It is well-known (but technically rather involved, {\it e.g.} \cite{DS}) how to construct quasimodes as well as the so-called interaction matrix to arbitrary order in $\kappa$. If all these quasimodes are spacially separated (which is the case if the multiplicities are all $i$), then tunnel effect estimates imply that the interaction matrix elements are of the order of $e^{-\frac{c}{\kappa}}$. {This can potentially lead to the splitting of eigenvalues of order $e^{-\frac{c}{\kappa}}$. In particular, the kernel can be lifted. As already pointed out based on \eqref{eq-ChiralGenSL}, the spectrum of the spectral localizer is always reflection symmetric. Hence eigenvalues can leave the kernel in pairs. This is indeed permitted because the total topological charge of the singular points vanishes, see Section~\ref{sec-TopCharge} below. Generically, all eigenvalues of $L_\kappa$ will then leave the kernel as $\kappa$ increases and form symmetric pairs. All of this can readily be verified numerically, see \cite{SSt1}.}

\subsection{Weyl point count in disordered semimetals}
\label{sec-PerturbRand}

Theorem~\ref{theo-SemiclassicsIntro} considers the spectral localizer of an ideal semimetal, hence a particular periodic system. It shows that the approximate kernel of its spectral localizer is separated from the rest of the spectrum. The aim of this short and non-rigorous section is to analyze that the dependence of the low-lying eigenvalues $\nu_i(\lambda)$, $i=1,\ldots,I$, of the spectral localizer $L_{\kappa}$ on a weak random matrix potential $\lambda V$. For sake of concreteness, we will focus on the case of odd dimension $d$ and suppose that the random term is of the diagonal form
$$
V
\;=\;
\sum_{n\in\ZM^d}
V_n\,|n\rangle\langle n|
\;,
$$
with i.i.d. and centered random matrices $V_n\in\CM^{N\times N}$. Multiplied with a coupling constant $\lambda\geq 0$ this is added to the periodic semimetal Hamiltonian $H$ and consequently also leads to a random spectral localizer
$$
L_\kappa(\lambda)
\;=\;
\begin{pmatrix}
-(H+\lambda V) & \kappa\, \widehat{D} \\
\kappa\, \widehat{D} & H+\lambda V  
\end{pmatrix}
\;=\;
L_\kappa
\;+\;\lambda
\begin{pmatrix}
-V  & 0\\
0 & V 
\end{pmatrix}
\;.
$$
The spread of the approximate kernel will be estimated in first order perturbation theory (in $\lambda$) by studying the matrix elements of the perturbation $\binom{-V\;0}{\;0 \;\;V}$ on the $I$-dimensional subspace of the approximate kernel which is spanned by the quasi-modes constructed in the proof of Theorem~\ref{theo-SemiclassicsIntro}. Suppose they are given by a set $\psi_1,\ldots,\psi_{I}$ of the orthonormal vectors. Concretely, they are given by the inverse Fourier transform $\Ff^*$ of the vectors defined in \eqref{eq-ApproxKernelVec}. Setting $\Psi=(\psi_1,\ldots,\psi_{I})$, this leads to the $I\times I$ interaction matrix
$$
\Vv
\;=\;
\langle\Psi|
\begin{pmatrix}
- V & 0  \\
0 & V 
\end{pmatrix}
|\Psi\rangle
\;.
$$
Let us stress that the random perturbation is like a random mass term and hence, independent of its sign, tends to move the low-lying eigenvalues away from zero. It will be argued now (as already stressed, without a rigorous proof) that the $I$ eigenvalues $\nu_1(\lambda),\ldots,\nu_I(\lambda)$ in the approximate kernel typically behave like
\begin{equation}
\label{eq-EigenValRandomSplit}
\nu_i(\lambda)\,-\,\nu_i(0)\;=\;\Oo(\lambda \kappa^{\frac{d}{4}})
\;.
\end{equation}
This indicates that there is a rather weak dependence of the approximate kernel on the random potential, a fact that was already observed in the numerical study \cite{SSt1}. Hence the spectral localizer can safely be used to detect Weyl points even in a weakly disordered system, or inversely, one can use the approximate kernel dimension to define the number of Weyl points in such a system.

\vspace{.2cm}

Let us now give some support for \eqref{eq-EigenValRandomSplit}. While intuitively based on the semiclassical picture already described in Section~\ref{sec-RoughSemiclassicalPicture}, we will recourse to some of the technical elements of the proof given in Section~\ref{sec-SemiClassicalLoc}. According to \eqref{eq-ApproxKernelVec} the quasimodes are given {by} $\psi_i=\Ff^* \chi_i^\delta W_i^* 0\oplus\phi_{\kappa,i}$  with $\chi_i^\delta$ a smooth indicator function on a neighborhood of $k^*_i$ and $\phi_{\kappa,i}$ being the Gaussian fundamental given in \eqref{eq-GaussState} (strictly speaking shifted, but this is irrelevant for the following rough argument). Now for $\delta>\kappa^{\frac{1}{2}}$, the factor  $\chi_i^\delta$ is negligible and $W$ is approximately constant and equal to the unitary matrix $W_{k^*_i}$ (for $\kappa$ small) so that $\psi_i\approx\Ff^* W_{k^*_i}\phi_{\kappa,i}$. Hence the matrix entries of $\Vv$ are approximately given by
$$
\Vv_{i,j}
\;\approx\;
\langle \phi_{\kappa,s_i}|
W_{k^*_i}^* \Ff \begin{pmatrix}
0 & V  \\
V & 0
\end{pmatrix}\Ff^*
W_{k^*_j}
|\phi_{\kappa,s_j}\rangle
\;.
$$
For further simplification, let us suppose that the $d$ slopes $s_j$ are of constant modulus $|s|$ (even for all $j$) and that the matrices $W_{k^*_i}$ merely introduce phase factors that can be neglected. The Fourier transforms of  the matrix entries can very roughly be estimated to be of the size
$$
\Vv_{i,j}
\;\approx\;
\sum_{n\in\ZM^d} V_n
\big(\pi \tfrac{|s|}{\kappa}\big)^{-\frac{d}{2}} e^{-\frac{\kappa}{ |s|} \,|n|^2}
\;\approx\;
\big(\pi \tfrac{|s|}{\kappa}\big)^{-\frac{d}{2}}
\sum_{|n|\leq (\frac{|s|}{\kappa})^{\frac{1}{2}}} V_n
\;\approx\;
\kappa^{\frac{d}{4}}\,\Oo(1)
\;,
$$
where the random term is order $1$ due to the central limit theorem. This indicates that \eqref{eq-EigenValRandomSplit} should hold.

\subsection{Topological charge and the Fermion doubling theorem} 
\label{sec-TopCharge}

Up to now, nothing concrete has been said about the topological charge of the singular points of an ideal semimetal. This section reviews some standard facts in this respect \cite{ArmitageEtAl,Van,MT}. Let us focus on $d\geq 3$ odd. Then for each singular point $k^*_i$ the associated topological charge is
$$
c_i^*
\;=\;
\Ch_{d-1}(Q,\partial B_\epsilon(k^*_i))
\;,
$$
where $Q_k=\sgn(H_k)$ and $B_\epsilon(k^*_i)$ is a ball of sufficiently small radius $\epsilon>0$. 
Often $c_i^*$ is also called the chirality of the Weyl point $k^*_i$. If the singular point $k^*_i$ is simple in the sense that there is merely one Weyl Hamiltonian as a summand  in \eqref{eq-WeylPoint}, then
$$
c_i^*
\;=\;
(-1)^{\frac{d+1}{2}}\,{\sgn(\det(B_i))}
\;.
$$
This is the result of  a rather standard computation, see {\it e.g.} \cite{CSB} where the normalization of the Chern number differs by $(-1)^{\frac{d+1}{2}}$ and formulas are written out for the projection $\frac{1}{2}(\one-Q)$ instead of the symmetry $Q$. Let us note that for twice degenerate Weyl point, typically the two chiralities are opposite so that $c^*_i=0$. Such a point is then called a Dirac point. In dimension $d=3$, it is generic for an inversion symmetric semimetal with odd time-reversal symmetry (that is, with half-integer spin). For the remaining odd number $d=1$, the topological charge is defined by
\begin{equation}
\label{eq-Chern0}
c_i^*
\;=\;
\frac{1}{2}\big(
\Sig(Q_{k^*_i+\epsilon})\,-\,\Sig(Q_{k^*_i-\epsilon})
\big)
\;,
\end{equation}
where the signature $\Sig(Q)$ of a selfadjoint matrix is the number of positive eigenvalues minus the number of negative eigenvalues of $Q$. Note that this coincides with the above formula in terms of {$\sgn(\det(B_i))$}. {One important result on the charges is the following sum rule.}

{
\begin{proposition}[Nielsen-Ninomiya fermion doubling theorem \cite{NN}]
\label{prop-NN}
Let $H$ be an ideal Weyl semimetal in odd dimension $d$, so in particular with finite-dimensional fibers. Then
\begin{equation}
\label{eq-NN}
\sum_{i=1}^{I} c_i^*
\;=\;0
\;.
\end{equation}
\end{proposition}
}

{A short index-theoretic proof is given in Section~\ref{sec-SemiclassicalCallias}. In particular, if all band-touching points have minimal degeneracy then there must be an even number of them. The charge cancellation of all band-touching points also remains valid in some more generality, {\sl e.g.} if the Hamiltonian is perturbed such that they are not all fixed to the same energy any more.}


\subsection{Topological mass terms} 
\label{sec-TopoMass}

In Section~\ref{sec-SemimetalTransition} it was noted that semimetals often arise as transition points between gapped topological insulators. In two dimensions this is an immediate consequence of the fact that gapping out a linear band-touching point causes a jump in the Chern number \cite{Bel95}. In this section, we argue similarly that starting out with a Dirac semimetal one can add gap opening topological {scalar} mass terms which lead to a transition between topological insulators with any prescribed jump of Chern numbers, merely bounded above by the sum $\sum_{i=1}^I|c^*_i|$ of the topological charges.  Again the focus will be on odd dimensions $d$, so let $H$ be an ideal Weyl semimetal. Then let us consider the spectral localizer $L_\kappa$ given in \eqref{eq-GenSpecLoc} and add a {scalar} mass term to it in a standard manner
\begin{equation}
\label{eq-GenSpecLocMass}
L_\kappa(m)
\;=\;
\begin{pmatrix}
-H & \kappa\,\widehat{D}-\imath m{Y} \\ \kappa\,\widehat{D}+\imath m{Y} & H
\end{pmatrix}
\;=\;
-\,H\otimes\sigma_3
\;+\;\kappa\,\widehat{D}\otimes \sigma_1
\;+\;m\,{Y}\otimes\sigma_2 
\;.
\end{equation}
Here $m\in\RM$ and ${Y}={Y}^*$ is a scalar operator in the sense that its Fourier transform $\Ff {Y}\Ff^*=\int^\oplus dk\,{Y}_k\,\one$ is a fiberwise multiplication with a real-valued function  $k\in\TM^d\mapsto {Y}_k$. We thus have  $[H,{Y}]=0$ and, moreover, will suppose that ${Y}_{k^*_i}\not=0$ for all $i=1,\ldots,I$. Then ${y^*_i=\sgn(Y_{k^*_i})\in\{-1,1\}}$ is well-defined. All these conditions can be satisfied if ${Y}$ is chosen to be a scalar polynomial in the shift operators. One example is to simply take ${Y}=\one$.  In order to understand $m{Y}$ also as a mass term of the Hamiltonian, it is useful to carry out a  unitary basis change $e$ as in \eqref{eq-PassageOddEven0}:
\begin{equation}
\label{eq-PassageOddEven}
e\,L_{\kappa}(m)\,e^*
\;=\;
\begin{pmatrix}
\kappa \widehat{D} & H-\imath m{Y} \\
H+\imath m {Y} & -\kappa \widehat{D}
\end{pmatrix}
\;.
\end{equation}
Hence setting $A=H+\imath m {Y}$ one obtains an effective chiral Hamiltonian $H_{\mbox{\rm\tiny eff}}(m)=\binom{0\;A^*}{A\;\;0}$ for which $eL_{\kappa}(m)e^*$ coincides with the odd spectral localizer $L^\od_\kappa$, {\it cf.} its definition \eqref{eq-oddSL}. Since the Weyl-points of $H$ are doubled $H_{\mbox{\rm\tiny eff}}(0)$ is a chiral Dirac-semimetal. Moreover, $H_{\mbox{\rm\tiny eff}}(m)$ is invertible because $\Ff A^*A\Ff=\Ff AA^*\Ff=\int^\oplus dk\,(H_k^2+m^2{Y}_k^2)$ is strictly positive because ${Y}_{k^*_i}\not=0$ by assumption. Consequently, for $m\not=0$ the effective Hamiltonian $H_{\mbox{\rm\tiny eff}}(m)$ describes an insulator and its $d$-th odd Chern number $\Ch_d(Q_{\mbox{\rm\tiny eff}}(m))\in\ZM$ is well-defined. Now $m\mapsto L_{\kappa}(m)$ is an analytic path of selfadjoint Fredholm operators and its spectral asymmetry is linked to the Chern number $\Ch_d(Q_{\mbox{\rm\tiny eff}}(m))$ by \eqref{eq-oddSLForm}. Hence
$$
\Ch_d(Q_{\mbox{\rm\tiny eff}}(m))
\,-\,
\Ch_d(Q_{\mbox{\rm\tiny eff}}(-m))
\;=\;
\frac{1}{2}\;
\SF\big(
m'\in[-m,m]\mapsto L_{\kappa}(m')
\big)
\;,
$$
similar as in \eqref{eq-ChernDiffSF}. The point of the next result is that one can choose the topological mass term such that the low-lying eigenvalues at $m=0$ as appearing in Theorem~\ref{theo-SemiclassicsIntro} each be moved to the left or right by suitably choosing the signs $b^*_i$. More precisely, if there is a degenerate singular point $k^*_i$, for each summand the chirality determines the direction in which the eigenvalue moves. In particular, for a twice-degenerate Dirac point with $c^*_i=1-1=0$, one eigenvalue will move to the right and one to the left, leading to no net spectral flow.

\begin{proposition}
\label{prop-ChernDiff}
Let $H=H^*$ be an ideal Weyl semimetal in odd dimension $d$. For a topological {scalar} mass term ${Y}$ constructed as above, one has for all $m>0$
$$
\Ch_d(Q_{\mbox{\rm\tiny eff}}(m))
\,-\,
\Ch_d(Q_{\mbox{\rm\tiny eff}}(-m))
\;=\;
(-1)^{\frac{d-1}{2}}\,\sum_{i=1}^I
{y}^*_ic^*_i
\;.
$$
\end{proposition}

The main relevant fact for the proof is that the difference of the two Chern numbers in the limit as $m\to 0$ is given by contributions stemming merely from the singular points. These contributions can be computed explicitly as two integrals that give $\pm \frac{1}{2}$. For even dimension, this is exactly as in the well-known fact that the "Chern numbers of a Dirac Hamiltonian are $\frac{1}{2}$", see {\it e.g.} \cite{Lei}. Actually, one computes integrals for the upper or lower band of a massive Dirac or Weyl Hamiltonian which look exactly as the integrals in \eqref{eq-EvenChern} and \eqref{eq-OddChern}. Those integrals do not have a topological interpretation on their own, though differences between them for different masses do and must therefore be integers \cite{Bal20}. Technical elements of the proof of Proposition~\ref{prop-ChernDiff} can be found in \cite{PSbook} based on the work of Golterman, Jansen and Kaplan \cite{GJK}, see also the more recent contributions by Bal \cite{Bal} and Drouot \cite{Dro}. We nevertheless provide full details in Section~\ref{sec-ChernIntegrals}. 

\vspace{.2cm}

Let us briefly indicate how to proceed similarly in even dimension $d$ for an ideal Dirac semimetal that is described by a chiral Hamiltonian $H=-\Gamma H\Gamma$ as in \eqref{eq-ChiralHam}. According to Theorem~\ref{theo-SemiclassicsIntro}, the count of the of Dirac points is then possible using the even spectral localizer $L^\ev_\kappa$. The mass term is then added as follows
\begin{equation}
\label{eq-EvenChiralMassAdd}
L_\kappa^\ev(m)
\;=\;
\begin{pmatrix}
-(H-m{Y}\Gamma ) & \kappa\,D_0^* \\ \kappa\,D_0 & H-m{Y}\Gamma 
\end{pmatrix}
\;,
\end{equation}
with ${Y}$ satisfying $[{Y},H]=0$ and not vanishing at the critical points. The associated gapped insulator Hamiltonian is then $H(m)=H-m{Y}\Gamma$. If $Q(m)=H(m)|H(m)|^{-1}$, it has even Chern number $\Ch_d(Q(m))$ for $m\not=0$. One can then formulate a statement analogous to Proposition~\ref{prop-ChernDiff}, the proof of which is based on the computation of the Chern integral of a mass-gapped Dirac operator carried out in Section~\ref{sec-ChernIntegrals}. The details are left to the reader.

\section{Results on Callias-type operators} 
\label{sec-Overview2}

\subsection{The Callias index theorem} 
\label{sec-ClassicalCallias}

As was already explained in Section~\ref{sec-RoughSemiclassicalPicture}, the approximate zero modes stem from local contributions given by the Weyl operator \eqref{eq-WeylIntro} in Fourier space and shifted into $k^*$, namely the approximate local toy model has a variable varying in $\RM^d$ rather than the torus $\TM^d$ and the spacial variable will therefore be denoted by $x$ so that the Weyl Hamiltonian (in Fourier space) reads
\begin{equation}
\label{eq-WeylHam}
x\in\RM^d\;\mapsto \;
H^W_x\;=\;
\sum_{j=1}^d {\langle x|Be_j\rangle\,\Gamma_j}
\;.
\end{equation}
{where $B$, $e_1,\ldots, e_d$ and $\Gamma_1,\ldots,\Gamma_d$ are as in Definition~\ref{def-IdealSemimet}.} Associated to the Weyl Hamiltonian is now a spectral localizer $L^W_\kappa$. Let us directly present the general definition of this spectral localizer for any given function $x\in\RM^d\mapsto H_x$ of hermitian matrices:
\begin{equation}
\label{eq-SpecLocRedefine}
L_\kappa
\;=\;
\begin{pmatrix}
0 & \kappa\,D\,-\,\imath H
\\
\kappa\,D\,+\,\imath H & 0
\end{pmatrix}
\;.
\end{equation}
One of the technical issues addressed below is to provide sufficient conditions on $H=(H_x)_{x\in\RM^d}$ for this to be a selfadjoint Fredholm operator. Note that, apart from the Fourier transform, $L_\kappa$ is indeed precisely the form of the spectral localizer given in \eqref{eq-SpecLocCrep}. It particular, just as in \eqref{eq-ChiralGenSL} it has a  chiral symmetry, which here reads
\begin{equation}
\label{eq-ChiralSym}
J\,L_\kappa\,J
\;=\;
-\,L_\kappa
\;,
\end{equation}
with $J=\one\otimes\sigma_3$. This relation is referred to as the supersymmetry of $L_\kappa$ \cite{CFKS}, or alternatively $L_\kappa$ is said to be $J$-hermitian in Krein space terminology. The off-diagonal entry 
\begin{equation}
\label{eq-CalliasDef}
D_{\kappa,H}
\;=\;
\kappa\,D\,+\,\imath H
\;,
\qquad
D\;=\;-\,\imath\,\sum_{j=1}^d \gamma_j \,\partial_j
\;.
\end{equation}
is called the Callias or Dirac-Schr\"odinger operator and in this context $H=(H_x)_{x\in\RM^d}$ is also called the Callias potential. The index of $D_{\kappa,H}$ is given by the signature of the quadratic form $J$ restricted to the finite dimensional subspace $\Ker(L_\kappa)$:
\begin{equation}
\label{eq-SusyInd}
\Ind(D_{\kappa,H})
\;=\;
\Sig\big(J|_{\Ker(L_\kappa)}\big)
\;.
\end{equation}
Hence the index of $D_{\kappa,H}$ is tightly connected to the kernel of $L_\kappa$. It will be shown in Section~\ref{sec-Weyl} by an explicit computation that the spectral localizer $L^W_\kappa$ associated to the Weyl operator $H^W$ has precisely one zero mode  and this therefore constitutes the index-theoretic input of the proof of Theorem~\ref{theo-SemiclassicsIntro}. The statement on the spectral localizer of the Weyl Hamiltonian is a special case of a well-known general index theorem \cite{Cal,GW} that we state next.

\begin{theorem}
\label{theo-CalliasFinite}
Let $d$ be odd. Suppose $x\in\RM^d\mapsto H_x\in\CM^{N\times N}$ is differentiable and such that for each $\mu \in \RM\setminus \{0\}$ the map $\xi \in C^\infty_c(\RM^d, \CM^N) \mapsto [\kappa D, H](H-\imath \mu)^{-1} \xi$ extends to a bounded operator on $L^2(\RM^d,\CM^N)$. Moreover, suppose that there exist constants $C>0$ and $R_c$ such that 
\begin{equation}
\label{eq-RadialBound}
(H_x)^2\,-\,\kappa\sum_{j=1}^d\|(\partial_j H)_x\|
\;\geq\;
C\,\one
\;,
\qquad
\forall\;|x|\geq R_c
\;.
\end{equation}
Then the associated Callias or Dirac-Schr\"odinger operator $D_{\kappa,H}$ is a Fredholm operator with index given by
$$
\Ind(D_{\kappa,H})
\;=\;
\Ch_{d-1}(Q^R,\partial B_R)
\;,
$$ 
where $R\geq R_c$ and $Q_x=\sgn(H_x)$.
\end{theorem}

Note that the Weyl Hamiltonian $H^W$ satisfies all the hypothesis of Theorem~\ref{theo-CalliasFinite}. Further note that if \eqref{eq-RadialBound} holds for some $\kappa$, then it also holds for all $\kappa'\in(0,\kappa]$ and hence the index $\Ind(D_{\kappa',H})$ is independent of $\kappa'\in(0,\kappa]$. Section~\ref{sec-ProofCallias} will provide a new proof of Callias index theorem, based on a homotopy argument and the explicit computation mentioned above.

\subsection{Spectral localization for the Callias operator} 
\label{sec-SemiclassicalCallias}

Theorem~\ref{theo-CalliasFinite} makes no assumption on the zero set $\Zz(H)=\{x\in\RM^d\,:\,\dim(\Ker(H_x))>0\}$. In particular, $\Zz(H)$ can consist of hypersurfaces (which in the generic case are of codimension $d-3$). If, however, the zero set consists only of singular points as in an ideal semimetal, it is again possible to make a more detailed statement about the low-lying spectrum of $L_\kappa$, similar to Theorem~\ref{theo-SemiclassicsIntro}:

\begin{theorem}
\label{theo-CalliasSemiclassics}
Let $d$ be odd and suppose that $x\in\RM^d\mapsto H_x\in\CM^{N\times N}$ satisfies the assumptions of {\rm Theorem~\ref{theo-CalliasFinite}} as well as items {\rm (i)} and {\rm (ii)} of {\rm Definition~\ref{def-IdealSemimet}}. 
Then there are constants $c$ and $C$ such that the spectrum of $L_\kappa$ in $[-c\kappa^{\frac{2}{3}},c\kappa^{\frac{2}{3}}]$ consists of the eigenvalues $\nu_{\kappa,j}$, $j=1,\ldots,\sum_{i=1}^I\mult^*_i$. There is no further spectrum in $[-C\kappa^{\frac{1}{2}},C\kappa^{\frac{1}{2}}]$, namely 
$$
\sigma(L_\kappa)\,\cap\,[-C\kappa^{\frac{1}{2}},C\kappa^{\frac{1}{2}}] 
\;=\;
\big\{\nu_{\kappa,j}\,:\,j=1,\ldots,\mbox{$\sum_{i=1}^I$}\mult^*_i\big\}
\;.
$$
The index of $D_{\kappa,H}$ can be written as a sum of local topological charges
\begin{equation}
\label{eq-RobSal}
\Ind(D_{\kappa,H})
\;=\;
\sum_{i=1}^I c^*_i
\;.
\end{equation}
\end{theorem}

The proof of Theorem~\ref{theo-CalliasSemiclassics} follows the same strategy as that of Theorem~\ref{theo-SemiclassicsIntro} described  in Section~\ref{sec-RoughSemiclassicalPicture}. Actually the only supplementary element is the semiclassical computation of the index by the supersymmetric formula \eqref{eq-SusyInd}. As explained in the next paragraphs, this can also be done on the torus and then the index on the l.h.s. of \eqref{eq-RobSal} vanishes so that one can deduce the Nielsen-Ninomiya relation \eqref{eq-NN}. The argument to Theorem~\ref{theo-CalliasSemiclassics} has numerous similarities with Witten's proof of the Morse index theorem \cite{Wit}, in the form exposed in Chapter 11 of  \cite{CFKS}. More specifically, the role of the Morse function is played by the Callias potential satisfying the assumptions (i) and (ii). The Dirac operator corresponds to the exterior derivative in \cite{CFKS} so that the square in both cases is the Laplace-Beltrami operator. The sum of the Betti numbers, given by the nullity of an exterior product of the Laplace-Beltrami operator, is the equivalent of the index of the spectral localizer. On the other side, the counterpart of the topological charges are the Morse indices. The only difference is that here the operators are intrinsically matrix-valued so that one has to use a matrix-valued semiclassical localization procedure and that the local contributions are stemming from the spectral localizer associated to Weyl Hamiltonians and not merely harmonic wells. Of course, there are the similar correspondences to Patodi's proof of the Gauss-Bonnet theorem as described in Chapter 12 of  \cite{CFKS}.

\vspace{.2cm}

Based on Theorem~\ref{theo-CalliasSemiclassics}, generically the low-lying spectrum of $L_\kappa$ looks as follows: the kernel of $L_\kappa$ has dimension equal to $|\Ind(D_{\kappa,H})|$, and lies in positive subspace of $J$ if $\Ind(D_{\kappa,H})>0$ and in the negative one otherwise; the remainder of the spectrum comes in pairs $(-\nu,\nu)$. Only accidentally these eigenvalue pairs may merge into the kernel. If all $c^*_i$ have the same sign and $\mult^*_i=|c^*_i|$, then all low-lying eigenvalues lie in the kernel of $L_\kappa$. {Let us stress that this spectral picture of $L_\kappa$ only holds if the zero set is discrete. For example, there is more low-lying spectrum if the zero set is a curve.} To conclude this section, let us now provide an index-theoretic proof of the Nielsen-Ninomiya theorem for an odd-dimensional ideal Weyl semimetal.

\vspace{.2cm}

\noindent 
{{\bf Proof} of Proposition~\ref{prop-NN}.
From the spectral localization argument below it will follow that Theorem~\ref{theo-CalliasSemiclassics}, more precisely \eqref{eq-RobSal},  also holds for a differentiable function $k\in\TM^d\mapsto H_k$ on the torus. Then note that the Dirac operator $D$ on the torus (acting on  $L^2(\TM^d,\CM^{d'})$) has a compact resolvent and a vanishing index. Now the Callias potential is a relatively compact perturbation of $D$ so that $D_{\kappa,H}$ given by \eqref{eq-CalliasDef} has the same index, namely $\Ind(D_{\kappa,H})=0$. By Theorem~\ref{theo-CalliasSemiclassics} this implies \eqref{eq-NN}.
\hfill $\Box$
}

\subsection{{Callias index as sum of local contributions}} 
\label{sec-HigherSpecFlow}

This section is about a generalization of Theorem~\ref{theo-CalliasSemiclassics} to families $x\in\RM^d\mapsto H_x$ of selfadjoint Fredholm operators on a possibly infinite-dimensional Hilbert space. Let us begin by interpreting Theorem~\ref{theo-CalliasSemiclassics} and, in particular, the identity \eqref{eq-RobSal} in the case $d=1$. Recall that the topological charges $c^*_i$ in this case are given by the difference of the half-signatures, see \eqref{eq-Chern0}, which is exactly equal to the effective number of eigenvalues that passed through $0$ at the singular point $x^*_i$. Summing over all these eigenvalue crossings is the definition precisely the spectral flow \cite{APS,RS}:
$$
\SF(x\in\RM\mapsto H_x)
\;=\;
\sum_{i=1}^I c^*_i
\;.
$$
Hence \eqref{eq-RobSal} for $d=1$ can be restated as
\begin{equation}
\label{eq-RobSal2}
\Ind(D_{\kappa,H})
\;=\;
\SF(x\in\RM\mapsto H_x)
\;.
\end{equation}
For matrix-valued functions, this equality is a well-known statement (actually already contained as a special case in Callias' work \cite{Cal}). It is, however, known that the equality \eqref{eq-RobSal2} also holds for suitable paths $x\in\RM\mapsto H_x\in\FM_{\sa}(\Hh)$ of self-adjoint Fredholm operators on a Hilbert space $\Hh$. The case when $H_x$ is unbounded and has a compact resolvent goes back to Robbin and Salamon \cite{RoS}, while the case of bounded $H_x$ was apparently first proved in \cite{AM}. It is the object of this section to generalize this result to higher odd dimension $d$, an issue that was also already briefly discussed in Section~8.1 of \cite{SSt2}.

\begin{theorem}
	\label{theo-CalliasSemiclassicsInfDim}
	Let $d\geq 3$ be odd and suppose that $x\in\RM^d\mapsto H_x\in\FM_\sa(\Hh)$ is differentiable in the sense that $[D,H]$ extends to a bounded operator. If 
\begin{enumerate}

\item[{\rm (i)}] $H$ has a finite zero set $\Zz(H)=\{x\in\RM^d\,:\,\dim(\Ker(H_x ))>0\}$,

\item[{\rm (ii)}] for each zero $x^*_i\in\Zz(H)$ there exists some $c_i > 0 $ such that $|H_x| \geq c_i |x-x_{i}^*|$ holds in a neighborhood of $x_{i}^*$,

\item[{\rm (iii)}] $|H_x|$ is bounded from below outside some ball $B_R(0)$,
\end{enumerate} 
then for $\kappa$ sufficiently small
\begin{equation}
\label{eq-RobSalFred}
\Ind(D_{\kappa,H})
		\;=\;
		\sum_{x^*_i\in\Zz(H)} \Ch_{d-1}\big(P_{a_i}Q,\partial B_\delta(x^*_i)\big)
		\;,
\end{equation}
where  $\partial B_\delta(x^*_i)$ is the surface of a $d$-dimensional ball $B_\delta(x^*_i)$  of sufficiently small radius $\delta$ around $x^*_i$, and $a_i>0$ is sufficiently small so that $P_{a_i,x}=\chi(|H_x|<a_i)$ has a range of finite and constant dimension for all $x\in B_\delta(x^*_i)$. As $Q_x=\sgn(H_x)$, one can view $x\in \RM^d \mapsto P_{a_i,x} Q_x$ as a map into the finite-dimensional selfadjoint unitary matrices of locally constant dimension $\Tr(P_{a_i,x})$ so that the above Chern number is well-defined.
\end{theorem}

While the assumption of a finite zero set has also been made elsewhere in related contexts \cite{FLZ,Kub}, it is much more restrictive than needed.  The invariance properties of the l.h.s. in \eqref{eq-RobSalFred} indicate that the index can be computed as a sum of local contributions more generally for all families that can be brought into a standard form with isolated linear band-touchings via (stable) homotopy and compactly supported compact perturbations. Indeed, this is always possible for $d=1$ under the stated assumption, but the same is likely not the case in higher dimensions since the set of singular points is generically extended and hence the construction of the perturbation cannot be reduced to a local problem.

\vspace{.2cm}

The r.h.s.  of \eqref{eq-RobSalFred} is reminiscent of Phillips' approach to spectral flow \cite{Phi}, which also works using local finite-dimensional projectors. Indeed, the Callias index in the case above can be interpreted as an instance of noncommutative spectral flow (in the sense that it generalizes a similar $K$-theoretic pairing as the ordinary spectral flow, see \cite{KL, vD, SSt} which also goes back to ideas of \cite{Wahl2008}). This terminology of spectral flow may be justified by the fact that the index (in the limit $\kappa\to 0$) does only depend on the low-lying spectrum of the family $H$ and is trivial unless the spectral gap closes somewhere. Furthermore, if $H_x=\Gamma\cdot f(x)$ is a Dirac vector field the index and thus the r.h.s. are equal to the so-called joint spectral flow \cite{Kub} of the family $H$. An independent and more topological proof of Theorem~\ref{theo-CalliasSemiclassicsInfDim} as well as generalizations with less rigid assumptions on the zero set can also be derived more directly using known cohomological formulas for the Callias index \cite{SSt2}.

\subsection{Modifications for the case of even dimension $d$} 
\label{sec-IntroEven}

This section indicates what are the necessary changes if systems in even dimension $d$ are considered. First of all, Theorem~\ref{theo-SemiclassicsIntro} already states that for an ideal Dirac semimetal the multiplicity of the low-lying spectrum of the even spectral localizer is equal to the number of Dirac points (one can also use $L_\kappa$ which due to \eqref{eq-PairingEvenEven} simply has twice as many low-lying eigenvalues). The proof is the same provided that the facts about the spectral localizer of the Weyl operator (discussed in Section~\ref{sec-Weyl}) are replaced by those about the spectral localizer of the Dirac operator (studied in Section~\ref{sec-SpecDocDiracPoints}). No further details are given. 

\vspace{.2cm}

Here the focus is rather on an index theorem similar to Theorem~\ref{theo-CalliasSemiclassics} but in even dimension. The statement will involve an even Callias operator that was already used in \cite{GH,FH,SSt2}. For its construction, let us first recall that in even dimension the Dirac operator $D$ {anti-commutes with its chirality operator $\gamma_0=(-\imath)^{d/2} \gamma_1\cdots \gamma_d$. In this convention, $(\gamma_1,...,\gamma_d,\gamma_0)$ generate a left-handed representation of $\CM_{d+1}$. In the grading $\CM^{d'} = \Sigma_+ \oplus \Sigma_-$ given by positive and negative subspaces of $\gamma_0$, the Dirac operator is off-diagonal with off-diagonal entries given by $D_0: H^1(\RM^d)\otimes \Sigma_+ \to L^2(\RM^d)\otimes \Sigma_-$ and the adjoint $D_0^*$, just as in \eqref{eq-DiracEvenRep}. Suppose that the Callias potential $x\in\RM^d\mapsto H_x\in\FM_\sa(\Hh)$ is chiral (or supersymmetric), namely there exists a self-adjoint unitary $\Gamma_0$ on $\Hh$ such that $\Gamma_0 H_x\Gamma_0 =-H_x$}. In the grading of $\Gamma_0$, namely $\Hh=\Hh_+\oplus \Hh_-$, $H$ is then off-diagonal with off-diagonal entries $A$ and $A^*$ as in \eqref{eq-ChiralHam} which can be considered functions $x\in\RM^d\mapsto A_x \in \Bb(\Hh_+,\Hh_-)$. The set of selfadjoint Fredholm operators {which anti-commute with $\Gamma_0$ is denoted by $\FM_{\ssa}(\Hh, \Gamma_0)$}. {Since $\Gamma_0$ commutes with $D_0$ the symmetry $J= \gamma_0 \otimes \Gamma_0$ anti-commutes with both $D$ and $H$. This results in a decomposition 
$$
\CM^{d'}\otimes \Hh 
\;=\; 
\left((\Sigma_+\otimes \Hh_+) \oplus (\Sigma_- \otimes \Hh_-)\right) \oplus \left((\Sigma_+\otimes \Hh_-) \oplus (\Sigma_- \otimes \Hh_+)\right)
$$ 
w.r.t. which the even spectral localizer $L^{\mathrm{ev}}_\kappa = \kappa D - \gamma_0 H$ is off-diagonal 
\begin{equation}
	\label{eq-EvenSpecLocOp}
	L^\ev_\kappa
	\;=\;
	\begin{pmatrix}
		0 & (D^\ev_{\kappa,H})^*
		\\
		D^\ev_{\kappa,H } & 0
	\end{pmatrix}
	\;.
\end{equation}
with off-diagonal component
\begin{equation}
\label{eq-EvenCalliasOp}
D^\ev_{\kappa,H}
\;=\;
\begin{pmatrix}
-\one_{\Sigma_+} \otimes A  & D_0^* \otimes \one_{\Hh_-}  \\ D_0 \otimes \one_{\Hh_+} & \one_{\Sigma_-}\otimes A^* 
\end{pmatrix}.
\end{equation}
The trivial tensor products with identity operators will be suppressed in the following (note that one then recovers the expression on the r.h.s. of \eqref{eq-SpecLocSSeven} for the even localizer).} The square of the localizer is again a matrix-valued Schr\"odinger operator:
\begin{equation}
\label{eq-SquareL'}
(L^\ev_\kappa)^2
\;=\;
-\kappa^2\,\Delta
\;+\;
\begin{pmatrix}
A^*A & \kappa [A,D_0]^* & 0 & 0 
\\
\kappa [A,D_0] & AA^* & 0 & 0
\\
0 & 0 & A^*A & \kappa [A^*,D_0]^*
\\
0 & 0 & \kappa [A^*,D_0] & AA^*
\end{pmatrix}
\;.
\end{equation}
If the hypothesis of Theorems~\ref{theo-CalliasFinite} and \ref{theo-CalliasSemiclassicsInfDim} hold, one can show  that $L^\ev_\kappa$ is a selfadjoint Fredholm operator (the arguments given in Section~\ref{sec-FredholmFinite} apply directly). Now the even-dimensional equivalent of Callias index theorem (Theorem~\ref{theo-CalliasFinite}) is the following:

\begin{theorem}
\label{theo-CalliasFiniteEven}
Let $d$ be even. Suppose $x\in\RM^d\mapsto H_x\in\CM^{N\times N}$ is differentiable and {anti-commutes with a self-adjoint unitary matrix $\Gamma_0$}. Further suppose  that for each $\mu \in \RM\setminus \{0\}$ the map $\xi \in C^\infty_c(\RM^d, \CM^N) \mapsto [\kappa D, H](H-\imath \mu)^{-1} \xi$ extends to a bounded operator on $L^2(\RM^d,\CM^N)$. Moreover, suppose that there exist constants $C>0$ and $R_c$ such that \eqref{eq-RadialBound} holds. Then the associated even Callias operator $D^\ev_{\kappa,H}$ is a Fredholm operator with index given by
$$
\Ind(D^\ev_{\kappa,H})
\;=\;
\Ch_{d-1}(Q^R,\partial B_R)
\;,
$$ 
where $R\geq R_c$ and $Q_x=\sgn(H_x)$.
\end{theorem}

{Finally let us note that also the equivalent of Theorem~\ref{theo-CalliasSemiclassicsInfDim} holds, if on the r.h.s. of \eqref{eq-RobSalFred} the odd $(d-1)$-th Chern numbers as defined in \eqref{eq-OddChern} are used.}

\begin{theorem}
\label{theo-CalliasEven}
Let $d\geq 2$ be even and suppose that {for some self-adjoint unitary $\Gamma_0$ on a Hilbert space $\Hh$ there is a map $x\in\RM^d\mapsto H_x\in\FM_\ssa(\Hh, \Gamma_0)$} that is differentiable and has a finite zero set $\Zz(H)$ such that for each zero $x^*_i\in\Zz(H)$ there exists some $c_i > 0 $ such that $|H_x| \geq c_i |x-x_{i}^*|$ holds in a neighborhood of $x_{i}^*$. Furthermore, assume that for each $\mu \in \RM\setminus \{0\}$ the map $\xi \in C^\infty_c(\RM^d, \Hh) \mapsto [\kappa D, H](H-\imath \mu)^{-1} \xi$ extends to a bounded operator on $L^2(\RM^d,\Hh)$. 
Then $D^\ev_{\kappa,H}$ is a Fredholm operator {for small enough $\kappa$} with index given by
\begin{equation}
\label{eq-RobSalEven}
\Ind(D^\ev_{\kappa,H})
\;=\;
{\sum_{x^*_i\in\Zz(H)} \Ch_{d-1}\big(P_{a_i}Q,\partial B_\delta(x^*_i)\big)}
\;,
\end{equation}
{where the r.h.s. is understood as in \eqref{eq-RobSalFred}.}
\end{theorem}

\section{Index computations}
\label{sec-ProofCallias}

The first object of this section is to provide a new short proof of the classical form of the Callias index theorem as stated in Theorem~\ref{theo-CalliasFinite}. Section~\ref{sec-FredholmFinite} first verifies the Fredholm property of the Callias operator, actually directly in a form that also covers the case with infinite dimensional fibers described in Section~\ref{sec-HigherSpecFlow}. Then in Section~\ref{sec-ReductionWeyl} follows a stable homotopy argument  deforming $H$ into a direct sum of Weyl Hamiltonians. The next Section~\ref{sec-Weyl} contains the computation of the index for the Weyl Hamiltonian. As already stressed above, this also constitutes an important element for the proof of Theorem~\ref{theo-SemiclassicsIntro}. Finally Section~\ref{sec-SpecDocDiracPoints} provides the necessary modifications for the even-dimensional analogues. 

\subsection{Fredholm property of the Callias operator}
\label{sec-FredholmFinite}

For sake of completeness, this short section indicates how the selfadjointness of the spectral localizer $L_\kappa$ as well as its Fredholm property (and thus the Fredholm property of the Callias operator $D_{\kappa,H}$) can be verified under the hypothesis stated in Theorems~\ref{theo-CalliasFinite} and \ref{theo-CalliasSemiclassics}. This is hardly novel  and actually essentially contained in earlier work \cite{KL0,vD,SSt2}.  Let us first verify that $L_\kappa$ is a selfadjoint operator, or equivalently that $D_{\kappa,H}^*=D_{\kappa,-H}$ with common domain $\mathrm{Dom}(D)\cap\mathrm{Dom}(H)$. If $x\in\RM^d\mapsto H_x$ is uniformly bounded, this directly follows from the Kato-Rellich theorem. To also deal with unbounded functions $x\mapsto H_x$ (which one has, {\it e.g.}, for the Weyl Hamiltonian), we appeal to \cite[Proposition 7.7]{KL0} which applies to the sum of odd self-adjoint operators with relatively bounded anti-commutator. Indeed, the hypotheses in that general result are satisfied by assumption as $x\mapsto H_x$ is differentiable and the map $\xi \in C^\infty_c(\RM^d, \Hh) \mapsto [\kappa D, H](H-\imath \mu)^{-1} \xi$ extends to a bounded operator for all $\mu \in \RM\setminus \{0\}$. Let us now give the well-known argument that the condition \eqref{eq-RadialBound} implies the Fredholm property of $L_\kappa$ and thus also the Callias operator $D_{\kappa,H}$. The square of $L_\kappa$ is given by
\begin{equation}
\label{eq-LSquare2}
(L_\kappa)^2
\;=\;
-\kappa^2\,\Delta\otimes \one
\;+\;
\begin{pmatrix}
(H_x)^2\;+\;\kappa\,(\gamma\cdot\partial H_x) & 0 \\
0 & (H_x)^2\;-\;\kappa\,(\gamma\cdot\partial H_x)
\end{pmatrix}
\;,
\end{equation}
where $\Delta$ is the $d$-dimensional Laplacian and $\gamma\cdot\partial=\sum_{j=1}^d\gamma_j\partial_j$. Hence the Fredholm property  is equivalent to a strictly positive lower bound on the essential spectrum of $(L_\kappa)^2$ which is a matrix-valued Schr\"odinger operator. The hypothesis \eqref{eq-RadialBound} implies that  the matrix potential is strictly positive outside a compact subset. It is well known that multiplication operators by compactly supported functions are relatively compact w.r.t. the Laplacian. Therefore Weyl's criterion on the essential spectrum (Section XIII.4 in \cite{RS}) shows that the essential spectrum of $(L_\kappa)^2$ is bounded away from $0$.

\subsection{Reduction to the case of the Weyl operator}
\label{sec-ReductionWeyl}

This section completes  the proof of Theorem~\ref{theo-CalliasFinite} by reducing it to the special case when $H$ is given by a direct sum of copies of the Weyl-Hamiltonian \eqref{eq-WeylHam}.  The associated selfadjoint unitary $Q^W=H^W|H^W|^{-1}$ on $\partial B_R$ is known to have a Chern number equal to 
\begin{equation}
\label{eq-WeylChern}
\Ch_{d-1}(Q^W,\partial B_R)
\;=\;
(-1)^{\frac{d+1}{2}}\,\sgn({\det(B)})
\;.
\end{equation}
{For the verification of this identity, it is helpful to first use the singular value decomposition of $B$, modified by a sign:
\begin{equation}
\label{eq-SingVal}
B
\;=\;
V_1\,\diag(b_1,\ldots,b_d)\,V_2
\;,
\end{equation}
where $V_1,V_2\in \mbox{\rm SO}(d)$ and $b_1,\dots,b_d$ are the singular values of $B$ modified by $1$ sign such that $\sgn(\det(B))=\prod_{j=1}^d\sgn(b_j)$. Then $\Gamma'=V_1^*\Gamma$ is another left-handed irreducible representation of $\CM_d$ and with $x'=V_2x$ one has 
\begin{equation}
\label{eq-WeylDiagonal}
H^W_x\;=\;
\sum_{j=1}^dx'_j b_j\Gamma_j'
\end{equation}
For the Weyl Hamiltonian in the latter form, the computation of the Chern number is}  a rather standard computation, see {\it e.g.} \cite{CSB} where the normalization of the Chern number differs by $(-1)^{\frac{d+1}{2}}$ and formulas are written out for the projection $\frac{1}{2}(\one-Q)$ instead of the symmetry $Q$. 

\begin{lemma}
\label{lem-reduce}
It is sufficient to prove {\rm Theorem~\ref{theo-CalliasFinite}} for the Weyl Hamiltonian $H^W$ as in \eqref{eq-WeylDiagonal}.
\end{lemma}

\noindent {\bf Proof.} 
For the reduction let us first deform a general selfadjoint multiplication operator $H$ satisfying the conditions of Section~
\ref{sec-FredholmFinite} into a direct sum of Weyl Hamiltonians on the inside of $B_{2R}(0)$. This is achieved in the framework of stable homotopy in the space $C(\SM^{d-1},\CM^{N\times N})$ of continuous vector-valued functions on the even dimensional sphere $\SM^{d-1}$. The $K_0$-group of homotopy classes of selfadjoint unitaries $Q$ (or alternatively projections $P=\frac{1}{2}(Q-1)$) with coefficients in $C(\SM^{d-1})$ is $\ZM\oplus\ZM$, with one $\ZM$ being the dimension of $P$ and the other the even Chern number given by \eqref{eq-EvenChern}. It is therefore not difficult to see using standard methods ({\it e.g.} \cite{WO}) that if $Q$ is a given selfadjoint unitary with Chern number $n=\Ch_{d-1}(Q)$, then there are $N,N',N''$ such that $Q\oplus \one_{N'}$ is homotopic to {$(Q^W)^{\otimes \lvert n\rvert}\oplus\one_{N''}$}  inside the set of selfadjoint unitaries in $C(\SM^{d-1},\CM^{N\times N})$. Here the orientation of the Weyl point $\sgn(s)$ can and must be chosen such that \begin{equation}
\label{eq:calliasfiniterhsweyl}
\Ch_{d-1}(Q)=\,\Ch_{d-1}((Q^W)^{\otimes \lvert n\rvert})=|n|\,\Ch_{d-1}(Q^W).\end{equation}
By rescaling space to achieve $R_c< \frac{1}{2}$ we can assume that $|H_x|$ is bounded from below uniformly for all $r >\frac{1}{2}$. For each such $r > \frac{1}{2}$ choose, retracting to unitaries and enlarging the size of the matrices if necessary, a selfadjoint pointwise invertible path $t\in [0,1]\mapsto Q_r(t)$ connecting $Q_r(0) = H|_{r\SM^{d-1}} \in C(\SM^{d-1}, \CM^{N\times N})$ with $Q_r(1)=Q^W \otimes \one_{|n|}$. By a standard smoothing argument the paths can be chosen two times differentiable with respect to $r$. For $r\in [0,\frac{1}{2}]$ we also set $Q_r(0)=  H|_{r\SM^{d-1}}$ for notational convenience.

\vspace{.1cm}

Let now $\eta \in C^\infty_c(\RM^d, [0,1])$ be a function equal to $1$ for $|x| \in (\frac{6}{8},\frac{10}{8})$ and vanishing for $|x| \notin (\frac{5}{8},\frac{11}{8})$. Then the path
$$
t \in [0,1] 
\;\mapsto\; 
Q_{|x|}(\eta(x)t)
$$ 
is norm-continuous uniformly with respect to $x$ and hence gives rise to a bounded norm-continuous path between $H$ for $t=0$ and some potential $\tilde{H}$ at $t=1$. One has $\tilde{H}_x=Q^W_x$ for all $|x| \in (\frac{6}{8},\frac{10}{8})$, $\tilde{H}_x=H_x$ for all $|x| \notin (\frac{5}{8},\frac{11}{8})$ and by construction $\tilde{H}_x$ is still invertible for all $|x| > \frac{1}{2}$.

\vspace{.1cm}

Finally, let $\chi$ be a function that is equal to $1$ for $|x|\in [0,1]$ and vanishing for $|x|\notin [0,\frac{9}{8}]$; and let $g$ be a function equal to $\mathrm{id}$ on $[0,\frac{6}{8}]$ and equal to $1$ on $[\frac{7}{8},1]$.
Then the path
$$t\in [0,1]\mapsto (1-\chi(x))\tilde{H}_x + (1-t) \chi(x) \tilde{H}_x + t\chi(x) g(|x|) Q^W_x\otimes \one_{|n|}$$
carries $\tilde{H}$ into a potential $\hat{H}$ that is equal to $H^W_x\otimes \one_{|n|}\oplus \one_{N''}$ within the ball of radius $\frac{6}{8}$ and invertible outside (since the right endpoint coincides with $g(|x|)Q^W_x\otimes \one_{|n|} \oplus \one_{N''}$ in $[0,1]$ and equals $\tilde{H}_x$ outside).

\vspace{.1cm}

All together this provides a stable homotopy from $H_x\oplus \one_{N'}$ to a potential which is invertible everywhere except at the isolated point $0\in \RM^d$ where it coincides with a direct sum of Weyl Hamiltonians $(H^W_x)^{\otimes n} \otimes\one_{|n|}\oplus\one_{N''}$ on a ball $B_{\delta}$ around $0$. Since the path can be implemented using bounded and norm-continuous perturbations it gives rise to a Riesz continuous deformation of $L_\kappa$. Along this homotopy, the Chern number also does not change. On the other hand, the extension to $N\times N$ matrices was by $\one_{N'}$ so that $\Ind(D_{\kappa,H})=\Ind(D_{\kappa,H\oplus \one_{N'}})$, and then along the above homotopy of Fredholm operators $\lambda\mapsto D_{\kappa,H(\lambda)}$ because the Fredholm property is conserved throughout. 

\vspace{.1cm}

With this preparation we can now further reduce the index computation to an actual Weyl-Hamiltonian. We form the block Hamiltonian $H^{(2)}=\hat{H} \oplus (-(H^W)^{\otimes \lvert n\rvert} \otimes \one_{|n|} \oplus \one_{N''})$, i.e. we add $|n|$ Weyl points with the opposite charge and complement the fiber to the same matrix size as $\hat{H}$. Forming the Callias type operator $D_{\kappa,H^{(2)}} = \kappa D\otimes \one + \imath \one \otimes H^{(2)}$ the homotopy invariance and additivity of the Fredholm index imply 
$$
\Ind(D_{\kappa,H^{(2)}})
\;=\; 
\Ind(D_{\kappa,H}) \,+\, |n| \Ind(D_{\kappa,-H^W})
\;.
$$
We will now show that the l.h.s. of this equation vanishes. Choose a smooth positive function $\chi_0$ supported on $B_{\delta/2}(0)$ and equal to $1$ on $B_{\delta/3}(0)$. Neither the Fredholm property nor the index is changed under such compactly supported perturbations  so we may also add a small mass term
$$
H_m^{(2)} 
\;=\;
\kappa D \otimes \one + \begin{pmatrix}
	\hat{H} & m \chi_0\otimes \one \\ m \chi_0\otimes \one & -(H^W)^{\otimes \lvert n\rvert} \otimes \one_{|n|} \oplus \one_{N''} 
\end{pmatrix}
$$
which combines the Weyl points of opposite charges and opens a gap for any $m\neq 0$ since the algebraic properties of the Clifford matrices imply 
$$
\begin{pmatrix}
	H^W_x & \epsilon \\ \epsilon & -H^W_x
\end{pmatrix}^2 
\;=\; 
(\epsilon^2 + |x|^2) \one
\;.
$$ 
By construction $H_m^{(2)}$ is therefore an invertible operator for arbitrarily small $m> 0$. Fixing some $m$ and then choosing $\tilde{\kappa}\leq \kappa$ small enough one concludes that $D_{ H_m^{(2)},\tilde{\kappa}}$ is also invertible since the right hand side of \eqref{eq-LSquare2} then becomes a strictly positive operator. Shrinking $\kappa$ does not change the index either since the Fredholm property is preserved, so we conclude  $\Ind(D_{\kappa,H^{(2)}})=\Ind(D_{\tilde{\kappa},H_m^{(2)}})=0$ and hence 
$$
\Ind(D_{\kappa,H}) 
\;=\; -|n|\, \Ind(D_{\kappa,-H^W}) 
\;=\; |n|\, \Ind(D_{\kappa,H^W})
$$
Comparing with \eqref{eq:calliasfiniterhsweyl} it is indeed sufficient to prove $\Ind(D_{\kappa,H^W})=\Ch_{d-1}(Q^W)$ to conclude Theorem~\ref{theo-CalliasFinite}.
\hfill $\Box$

\subsection{The spectral localizer of the Weyl operator}
\label{sec-Weyl}

{Recall that the Weyl-Hamiltonian (in Fourier space) is given by \eqref{eq-WeylHam} and then the associated spectral localizer $L^W_\kappa$ by \eqref{eq-SpecLocRedefine} with Dirac operator $D=-\imath\sum_{j=1}^d\gamma_j\partial_j$ as in \eqref{eq-CalliasDef}. Let us bring the Weyl operator into the standard form \eqref{eq-WeylDiagonal} by use of the modified singular value decomposition \eqref{eq-SingVal} of $B$. Using the associated unitary transformation $\Vv_1:L^2(\RM^d)\to L^2(\RM^d)$ given by $(\Vv_1\psi)(x)=\psi(V_1^*x)$, one has $\Vv_1 D\Vv_1^*=-\imath\sum_{j=1}^d\gamma'_j\partial_j$ where $\gamma'=V_1\gamma$ is another left-handed irreducible representation of $\CM_d$. Hence
$$
\Vv_1\,L^W_\kappa\,\Vv_1^*
\;=\;
\begin{pmatrix}
0 & \kappa\,D'-\imath H^{W'} \\ \kappa\,D'+\imath H^{W'} & 0
\end{pmatrix}
\;,
\qquad
D'\,=\,-\imath\sum_{j=1}^d\gamma'_j\partial_j
\;,
\;\;H^{W'}\,=\,\sum_{j=1}^dX_j b_j\Gamma_j'
\;,
$$
where $X_j$ denote the components of the (selfadjoint) position operator on  $L^2(\RM^d)$ {densely defined by $(X_j\psi)(x)=x_j\psi(x)$}. Note that $\gamma$ and $\Gamma$ are both left-handed irreducible representations and that they still commute. Hence the problem of computing the spectrum of $L_\kappa$ is reduced to the case where the Weyl operator is of the diagonal form with diagonal coefficients $b_1,\ldots,b_d$ given by the singular values of $B$, up to a sign such that $\prod_{j=1}^d\sgn(b_j)=\sgn(\det(B))$. In the following we will focus on this case and suppress all primes in the above equation, as well as the unitary $\Vv_1$.}

\vspace{.2cm}

Using the canonical commutation relations $\imath [\frac{1}{\imath}\partial_i,X_j]=\delta_{i,j}\,\one$, the square of the localizer can be computed explicitly: 
\begin{equation}
\label{eq-L^2MDef}
(L^W_\kappa)^2
\;=\;
\begin{pmatrix}
\sum_{j=1}^d (-\kappa^2\partial_j^2+{b}_j^2X_j^2)\,+\, \kappa M & 0
\\
0
&
\sum_{j=1}^d (-\kappa^2\partial_j^2+{b}_j^2X_j^2)\,
-\, \kappa M
\end{pmatrix}
\;,
\end{equation}
where the selfadjoint matrix $M$ is given by
\begin{equation}
\label{eq-MDef}
M
\;=\;
 \sum_{j=1}^d{b_j\,\gamma_j\otimes\Gamma_j}
\;.
\end{equation}
The spectrum of the sum of $d$ harmonic oscillators is $\{\kappa \sum_{j=1}^d|{b}_j|(2n_j+1)\,:\,n_j\geq 0\}$. Hence the lowest level is $\kappa|{b}|$ where $|b|=\sum_{j=1}^d|{b}_j|$. Now note that \eqref{eq-MDef} immediately implies the inequality $\|M\|\leq |{b}|$. In the sequel, it will be shown that the spectrum of $M$ contains either the value $|{b}|$ or $-|{b}|$ as a simple eigenvalue. This leads to a one-dimensional kernel of $L^W_\kappa$. Moreover, the first excited state of $(L^W_\kappa)^2$ is of order $\kappa$, so that the first excited state of $L^W_\kappa$ is of order $\sqrt{\kappa}$.

\vspace{.2cm}

\noindent {\bf Example} Let us first consider $d=1$ so that $d'=1$, $\gamma_1=1=\Gamma_1$ and $H^W={b}X$ where ${b}_1={b}$. Let $\phi_{\kappa,{b}}(x)=(\pi \frac{\kappa}{|{b}|})^{-\frac{1}{4}} e^{-\frac{|{b}|}{\kappa} \frac{x^2}{2}}$ be the square-normalized Gaussian. Then
$$
(\kappa D\,\pm\,\imath \,H^W)\phi_{\kappa,{b}}
\;=\;
-\imath(\kappa \partial\,\mp\, {b}X)\phi_{\kappa,{b}}
\;=\;
-\imath(-|s|\mp {b})X\phi_{\kappa,{b}}
\;,
$$
which implies that $\Ker(D_{\kappa,H^W})$ is spanned by $\phi_{\kappa,{b}}$ if ${b}<0$, while for ${b}>0$ the kernel of $D_{\kappa,H^W}$ is trivial, and vice versa for $D_{\kappa,H^W}^*$. Hence $\Ind(D_{\kappa,H^W})=-\sgn({b})$.
\hfill $\diamond$

\vspace{.2cm}

For the general analysis of $M$, let us first remark that $M$ is a linear combination of $d$ commuting symmetries $\gamma_1\otimes\Gamma_1,\ldots,\gamma_d\otimes\Gamma_d$. Hence  one can find a unitary basis change so that all these $d$ symmetries are simultaneously diagonal with $2^{d-1}$ signs on the diagonal. To get more information on the joint spectra, let us set
$$
\Ee^j_{\eta_j}
\;=\;
\Ker\big(\gamma_j\otimes\Gamma_j\,-\,\eta_j\,\one\big)
\;,
\qquad
\eta_j\in\{-1,1\}
\;,
$$
and for $\eta=(\eta_1,\ldots,\eta_d)\in\{-1,1\}^d$
$$
\Ff^d_\eta
\;=\;
\bigcap_{j=1}^d\,\Ee^j_{\eta_j}
\;.
$$
Then
\begin{equation}
\label{eq-MEigenvectors}
Mv\;=\;\eta\cdot {b}\,v
\;,
\qquad
\eta\cdot {b}=\sum_{j=1}^d\eta_j{b}_j
\;,
\;\;\;
v\in\Ff^d_\eta
\;.
\end{equation}
Therefore the multiplicities of eigenvalues of $M$ can be read off the following lemma. 

\begin{lemma}
\label{lem-JointEigenspaces}
Let $d$ be odd and $\eta\in\{-1,1\}^d$. Then 
$$
\dim(\Ff^d_\eta)
\;=\;
\left\{
\begin{array}{cc}
1 \;, & \mbox{ if } d=1\,{\rm mod}\,4 \mbox{ and an even number of }-1 \mbox{ in }\eta\;,
\\
1 \;, & \mbox{if } d=3\,{\rm mod}\,4 \mbox{ and an odd number of }-1 \mbox{ in }\eta\;,
\\
0\;, & \mbox{ otherwise }.
\end{array}
\right.
$$
\end{lemma}

\noindent {\bf Proof.} Let us first consider the case $d=3$. Up to unitary equivalence the matrices $\gamma_j\otimes \Gamma_j=\sigma_j\otimes\sigma_j$ are given by the three Pauli matrices. The eigenspace are explicitly computable, {\it e.g.} $\Ee^1_+={\rm span}\{\binom{1}{1}\otimes\binom{1}{1},\binom{1}{-1}\otimes\binom{1}{-1}\}$ and $\Ee^1_-={\rm span}\{\binom{1}{1}\otimes\binom{1}{-1},\binom{1}{1}\otimes\binom{1}{-1}\}$. Under an isomorphism $\CM^2\otimes\CM^2\cong \CM^4$ they become
$$
\Ee^1_+
\;=\;
{\rm span}
\left\{
\begin{pmatrix}
1 \\ 1 \\ 1\\ 1
\end{pmatrix}
,
\begin{pmatrix}
1 \\ -1 \\ -1\\ 1
\end{pmatrix}
\right\}
\;,
\qquad
\Ee^1_-
\;=\;
{\rm span}
\left\{
\begin{pmatrix}
1 \\ 1 \\ -1\\ -1
\end{pmatrix}
,
\begin{pmatrix}
-1 \\ -1 \\ 1\\ 1
\end{pmatrix}
\right\}
\;.
$$
Similarly
$$
\Ee^2_+
\;=\;
{\rm span}
\left\{
\begin{pmatrix}
1 \\ \imath \\ \imath \\ -1
\end{pmatrix}
,
\begin{pmatrix}
1 \\ -\imath \\ -\imath \\ -1
\end{pmatrix}
\right\}
\;,
\qquad
\Ee^2_-
\;=\;
{\rm span}
\left\{
\begin{pmatrix}
1 \\ \imath \\ -\imath \\ 1
\end{pmatrix}
,
\begin{pmatrix}
1 \\ -\imath \\ \imath \\ 1
\end{pmatrix}
\right\}
\;,
$$
and
$$
\Ee^3_+
\;=\;
{\rm span}
\left\{
\begin{pmatrix}
1 \\ 0 \\ 0\\ 0
\end{pmatrix}
,
\begin{pmatrix}
0 \\ 0 \\ 0 \\ 1
\end{pmatrix}
\right\}
\;,
\qquad
\Ee^3_-
\;=\;
{\rm span}
\left\{
\begin{pmatrix}
0 \\ 0 \\ 1\\ 0
\end{pmatrix}
,
\begin{pmatrix}
0 \\ 1 \\ 0\\ 0
\end{pmatrix}
\right\}
\;.
$$
With some care, one reads off that $\dim(\Ee^1_-\cap\Ee^2_-\cap\Ee^3_-)=1$ while $\dim(\Ee^1_+\cap\Ee^2_+\cap\Ee^3_+)=1$. Furthermore, 
$$
1
\;=\;
\dim(\Ee^1_+\cap\Ee^2_+\cap\Ee^3_-)
\;=\;
\dim(\Ee^1_+\cap\Ee^2_-\cap\Ee^3_+)
\;=\;
\dim(\Ee^1_-\cap\Ee^2_+\cap\Ee^3_+)
\;.
$$
Hence one has found all $4$ common eigenspaces and proved the claim for $d=3$. For the general case, one can proceed inductively from $d-2$ to $d$. Let $\gamma_1\otimes \Gamma_1,\ldots,\gamma_{d-2}\otimes \Gamma_{d-2}$ are $d-2$ be commuting symmetries built from two irreducible representations of the Clifford algebra $\CM_{d-2}$, and let $\Ff^{d-2}_{\eta'}$ for $\eta'\in\{-1,1\}^{d-2}$ be the joint eigenspaces. Then  $d$ commuting symmetries of the irreducible representations of $\CM_d$ are given by
$$
\gamma_1\otimes \Gamma_1\otimes\sigma_1\otimes\sigma_1
\;,\;
\ldots
\;,\;
\gamma_{d-2}\otimes \Gamma_{d-2}\otimes\sigma_1\otimes\sigma_1
\;,\;
\one\otimes\one\otimes\sigma_2\otimes\sigma_2
\;,\;
\one\otimes\one\otimes\sigma_3\otimes\sigma_3
\;.
$$
Then tensorizing $v\in\Ff^{d-2}_{\eta'}$ with the above $4$ common eigenvectors of $\sigma_j\otimes\sigma_j$, one obtains $4$ new eigenvectors of the $d$ symmetries with joint eigenvalues $\eta\in\{-1,1\}^d$ given by $(-\eta',-1,-1)$, $(-\eta',1,1)$, $(\eta',1,-1)$ and $(\eta',-1,1)$. Note that one of the first two leads to a new eigenvalue of largest modulus $\sum_{j=1}^d\eta_j$ and this is the only new value of  $\sum_{j=1}^d\eta_j$ added (these sums are precisely the eigenvalues of $M$ for the case ${b}_j=1$). Furthermore, if $\eta'$ has an even (odd) number of $+1$, then all $4$ possible $\eta$ have an odd (even) number of $+1$.
\hfill $\Box$

\vspace{.2cm}

\noindent {\bf Remarks} The above iterative procedure shows that for $\frac{d-1}{2}$ even one has a joint eigenvalue $\eta=(1,\ldots,1)$, but not $\eta=(-1,\ldots,-1)$, while for $\frac{d-1}{2}$ odd there is a joint eigenvalue $\eta=(-1,\ldots,-1)$, but not $\eta=(1,\ldots,1)$. From these extremal eigenvalues with eigenvector $w$, all other eigenvectors can be constructed from unitary raising and lowering operators
$$
R_{k,l}\;=\;\gamma_k\otimes \Gamma_l
\;,
\qquad
k,l\in\{1,\ldots,d\}\;,\;\;k\not=l
\;.
$$
If $v\in\Ff^d_\eta$, namely $\gamma_j\otimes\Gamma_j v=\eta_j v$, then
$$
\gamma_j\otimes\Gamma_j
\;
R_{k,l}v
\;=\;
\left\{
\begin{array}{cc}
\eta_j \,R_{k,l}v\;,
& j\not\in\{k,l\}\;,
\\
-\eta_j \,R_{k,l}v\;,
& j\in\{k,l\}\;,
\end{array}
\right.
$$
Thus each $R_{k,l}$ changes two of the joint eigenvalues and produces a new joint eigenvector. This can be done iteratively to construct all joint eigenstates from the extremal one $w$, by flipping two of the eigenvalues at each step. When flipping $2j$ of them, one has $\binom{d}{2j}$ choices. Hence, summing up all these possibilities,
$$
1\,+\,
\sum_{j=1}^{\frac{d-1}{2}}\binom{d}{2j}
\;=\;
1\,+\,
\sum_{j=1}^{\frac{d-1}{2}}
\left(
\binom{d-1}{2j}
\,+\,
\binom{d-1}{2j-1}
\right)
\;=\;
\sum_{k=0}^{d-1}
\binom{d-1}{k}
\;=\;
2^{d-1}
\;,
$$
showing that one indeed produces all eigenvectors in this manner.
\hfill $\diamond$

\begin{lemma}
\label{lem-OddMSpec}
Let $d$ be odd and set $\sgn({b})=\prod_{j=1}^d\sgn({b}_j)$. Then the spectrum of $M$ contains either $|{b}|$ or $-|{b}|$ as simple eigenvalue. The eigenvalue $|{b}|$ appears if and only if either $(\sgn({b})=1$ and $d=1\,{\rm mod}\,4)$ or $(\sgn({b})=-1$ and $d=3\,{\rm mod}\,4)$.
\end{lemma}

\noindent {\bf Proof.} This is based on \eqref{eq-MEigenvectors}. Let $\sgn({b})=1$, then an odd number of ${b}_j$ are positive and and even number of ${b}_j$ are negative. Then $\eta\cdot {b}=|{b}|$ for a suitable $\eta$ with an even number of $-1$. By Lemma~\ref{lem-JointEigenspaces}, a corresponding eigenvector only exists if $d=1\,{\rm mod}\,4)$. For the other cases one argues similarly.
\hfill $\Box$

\vspace{.2cm}

Now we can state and prove the main result of this section which together with the results of Section~\ref{sec-ReductionWeyl} completes the proof of Theorem~\ref{theo-CalliasFinite}.

\begin{proposition}
\label{prop-WeylSpecLocKernel}
For all $\kappa\in(0,\infty)$, $D_{\kappa,H^W}$ is a Fredholm operator and for all $R>0$
$$
\Ind(D_{\kappa,H^W})
\;=\;
(-1)^{\frac{d+1}{2}}\,\sgn({b})
{\;=\;
(-1)^{\frac{d+1}{2}}\,\sgn(\det(B))}
\;=\;
\Ch_{d-1}(Q^W,\partial B_R)
\;.
$$ 
\end{proposition}

\noindent {\bf Proof.} The index will be computed using \eqref{eq-SusyInd}. The kernel of $(L^W_\kappa)^2$ and thus $L^W_\kappa$ is constructed from the  tensor Gaussian product state
$$
\tilde{\phi}^d_{\kappa,{b}}\;=\;\tilde{\phi}_{\kappa,{b}_1}\otimes\cdots\otimes\tilde{\phi}_{\kappa,{b}_d}
\;,
$$
with $\tilde{\phi}_{\kappa,{b}}(x)=(\pi \frac{\kappa}{|{b}|})^{-\frac{1}{4}} e^{-\frac{|{b}|}{\kappa} \frac{x^2}{2}}$ as above. This kernel is always of dimension $1$. If either $(\sgn({b})=1$ and $d=1\,{\rm mod}\,4)$ or $(\sgn({b})=-1$ and $d=3\,{\rm mod}\,4)$, there is a non-zero vector $v$ such that $Mv=|{b}|v$ (see Lemma~\ref{lem-OddMSpec}). Hence by \eqref{eq-L^2MDef}
\begin{equation}
\label{eq-GaussState}
(L^W_\kappa)^2
{\phi}_{\kappa,{b}}
\;=\;
0\;,
\qquad
{\phi}_{\kappa,{b}}
\;=\;
\begin{pmatrix}
0 \\
\tilde{\phi}^d_{\kappa,{b}}\otimes v
\end{pmatrix}
\;=\;
0
\;,
\end{equation}
so that the signature is $\Sig(J|_{{\rm Ker}(L^W_\kappa)})=-1$. If on the other hand, $(\sgn({b})=1$ and $d=3\,{\rm mod}\,4)$ or $(\sgn({b})=-1$ and $d=1\,{\rm mod}\,4)$, one finds similarly $\Sig(J|_{{\rm Ker}(L^W_\kappa)})=1$ because the zero mode ${\phi}_{\kappa,{b}}$ then vanishes in the second component. Together this implies the first equality of the claim. That the second equality holds was already stated in Section~\ref{sec-ReductionWeyl}.
\hfill $\Box$

\vspace{.2cm}

\noindent {\bf Remark} Another perspective on the above computation is obtained by introducing the ``covariant derivatives''
$$
\nabla_j
\;=\;
-\,\imath\,\kappa\,\gamma_j\,\partial_j
\;+\;
\imath\,{b}_j\,\Gamma_j\,X_j
\;.
$$
They satisfy $\nabla_j^*\nabla_j=-\kappa^2\partial_j^2+x_j^2+\kappa {b}_j\gamma_j\Gamma_j$ as well as $\nabla_i\nabla_j=\nabla_j\nabla_i$ and $\nabla_i^*\nabla_j=\nabla_j\nabla_i^*$ for $i\not=j$. The kernel of $\nabla_j$ is spanned by Gaussians. As $\kappa D+\imath H^W=\sum_{j=1}^d\nabla_j$ one is led to study the joint kernel.
\hfill $\diamond$

\subsection{The spectral localizer of the Dirac operator}
\label{sec-SpecDocDiracPoints}

This section provides those supplementary elements for the proof of the even-dimensional case of Theorem~\ref{theo-SemiclassicsIntro} as well as Theorem~\ref{theo-CalliasEven} which are not identical to the proofs in the odd-dimensional case. This concerns essentially the computation of the spectrum of the spectral localizer of a Dirac operator (instead of the Weyl operator). We will also state an even-dimensional Callias-type index theorem. Hence the dimension $d$ is even and throughout we fix a self-adjoint unitary $\Gamma_0$ to impose the chirality condition $H=-\Gamma_0 H\Gamma_0$ \eqref{eq-ChiralHam}. As already explained in Section~\ref{sec-IntroEven}, this naturally leads to the even Callias operator $D^\ev_{\kappa,H}$ defined in \eqref{eq-EvenCalliasOp} as well as the associated even spectral localizer $L^\ev_\kappa$ given in \eqref{eq-EvenSpecLocOp}. As the hypothesis are the same as in Theorems~\ref{theo-CalliasFinite} and \ref{theo-CalliasSemiclassicsInfDim}, the arguments of Section~\ref{sec-FredholmFinite} directly apply and show that $L^\ev_\kappa$ is a selfadjoint Fredholm operator so that also $D^\ev_{\kappa,H}$ is a Fredholm operator. Furthermore, using the stable homotopies in the unitaries over an odd dimensional sphere (namely, in the group $K_1(C(\SM^{d-1}))\cong\ZM$) one can transpose the reasoning in Section~\ref{sec-ReductionWeyl} to show that it is sufficient to prove Theorem~\ref{theo-CalliasEven} {for the special case of a Dirac operator (in Fourier space) of the form
\begin{equation}
	H^D_{x}
	\;=\;
	\sum_{j=1}^d \langle x | B e_j \rangle \Gamma_j
	\;,
\end{equation}
with $\Gamma_1,...,\Gamma_d$ generators of an irreducible representation of $\CM_d$ for which $\Gamma_0 = (-\imath)^{d/2} \Gamma_1\cdots \Gamma_d$. Arguing as in the odd case, the Hamiltonian and spectral localizer are unitarily equivalent to a standard form, hence it is enough to consider the partially diagonalized form}
\begin{equation}
\label{eq-Dirac}
H^D_{x}
\;=\;
\sum_{j=1}^d {b}_{j}\Gamma_jx_j
\;,
\end{equation}
with non-vanishing coefficients ${b}_{1},\ldots,{b}_{d}$. It is a selfadjoint operator on $L^2(\RM^d,\CM^{d'})$ that anti-commutes with $\Gamma_0$, just as in \eqref{eq-ChiralHam}. {Due to unitary equivalence we may without loss of generality assume $\Gamma_j=\Gamma'_j\otimes\sigma_1$ for some left-handed irreducible representation $\Gamma_1',...,\Gamma_{d-1}'$ of $\CM_{d-1}$ and  $\Gamma_{d}=\one\otimes\sigma_2$ such that $\Gamma_{0}=\one\otimes\sigma_3$.} Then $A=\sum_{j=1}^{d-1}{b}_j\Gamma'_jx_j+\imath {b}_dx_d$ for the Dirac Hamiltonian.  Proposition~\ref{prop-DiracSpecLocKernel} below provides an index theorem for the Dirac Hamiltonian which parallels the statement for the Weyl Hamiltonian given in Proposition~\ref{prop-WeylSpecLocKernel}.  This then also proves Theorem~\ref{theo-CalliasEven}.

\vspace{.2cm}

We hence focus on the analysis of the Dirac operator \eqref{eq-Dirac}. Let $L^{\ev,D}_\kappa$ be the associated spectral localizer. According to \eqref{eq-SquareL'}, its square is given by
$$
(L^{\ev,D}_\kappa)^2
\,=\,
\big(-\kappa^2\,\Delta\,+\,\sum_{j=1}^d {b}_j^2x_j^2\big)\otimes \one
\,+\,\kappa\,\one\otimes K
\;,
$$
where
$$
K
\;=\;
\begin{pmatrix}
0 & -\imath (M-{b}_d) & 0 & 0 
\\
\imath (M-{b}_d) & 0 & 0 & 0
\\
0 & 0 &  0 & -\imath (M+{b}_d)
\\
0 & 0 & \imath (M+{b}_d) & 0
\end{pmatrix}
$$
with $M=\sum_{j=1}^{d-1}{b}_j \gamma'_j\Gamma'_j$ as in \eqref{eq-MDef}, with an odd number of summands as here $d-1$ is odd. Going into the fundamental of the harmonic oscillators as in Section~\ref{sec-Weyl}, one is hence interested to know whether $-|{b}|$ lies in the spectra of $K$. Note that it is of size $2^{d}\times 2^{d}$. 

\begin{lemma}
\label{lem-MEven}
Let $d$ be even. Then $-|s|$ is a simple eigenvalue of $K$. It lies in the upper $2\times 2$ block if and only if 
$(\sgn({b})=1$ and $d=0\,{\rm mod}\,4)$ or $(\sgn({b})=-1$ and $d=2\,{\rm mod}\,4)$.
\end{lemma}

\noindent {\bf Proof.}  Set ${b}'=({b}_1,\ldots,{b}_{d-1})$,  $\sgn({b}')=\prod_{j=1}^{d-1}\sgn({b}_j)$ and $|{b}|=\sum_{j=1}^{d-1}|{b}_j|$ as for ${b}$. Note that Lemma~\ref{lem-OddMSpec} applies to $M$, namely $M$ contains either $|{b}'|$ or $-|{b}'|$ as simple eigenvalue and the remainder of the spectrum is in $(-|{b}'|,|{b}'|)$. The eigenvalue $|{b}'|$ appears if and only if either $(\sgn({b}')=1$ and $d-1=1\,{\rm mod}\,4)$ or $(\sgn({b}')=-1$ and $d-1=3\,{\rm mod}\,4)$. If then $\sgn({b}_d)=-1$, it leads to an eigenvalue $-|{b}|$ in the upper block, while for $\sgn({b}_d)=1$, it leads to an eigenvalue $-|{b}|$ in the lower block. From this and analogous statements for an eigenvalue $-|{b}'|$ of $M$, one can check the claim.
\hfill $\Box$

\vspace{.2cm}

Given Lemma~\ref{lem-MEven}, one can now deduce an analogous statement to Proposition~\ref{prop-WeylSpecLocKernel} for the Dirac Hamiltonian, essentially by the same proof. The computation of the odd Chern number of a Dirac monopole is also given in \cite{CSB}.

\begin{proposition}
\label{prop-DiracSpecLocKernel}
For all $\kappa\in(0,\infty)$, $D^\ev_{\kappa,H^D}$ is a Fredholm operator and for all $R>0$
$$
\Ind(D^\ev_{\kappa,H^D})
\;=\;
(-1)^{\frac{d}{2}}\,\sgn({b})
{\;=\;
(-1)^{\frac{d}{2}}\,\sgn(\det(B))}
\;=\;
\Ch_{d-1}(Q^D,\partial B_R)
\;,
$$ 
where  $Q^D_x=\sgn(H^D_x)$.
\end{proposition}

Based on this result, the arguments of Section~\ref{sec-SemiClassicalLoc} transpose the even-dimensional case and also allow to prove Theorems~\ref{theo-CalliasFiniteEven} and \ref{theo-CalliasEven}. No further details are provided.

\section{Semiclassical spectral localization for $L_\kappa$}
\label{sec-SemiClassicalLoc}

This section presents the core of the semiclassical argument used to localize the spectrum $L_\kappa$. It will be spell out the case $x\in{\Mm^d}\mapsto H_x\in\FM_\sa(\Hh)$ where $d$ is odd, $x$ varies in a $d$-dimensional flat manifold $\Mm^d$ which is either {$\TM^d=\RM^d/\ZM^d\cong[0,1)^d$} or $\RM^d$ and $H_x$ acts on a possibly infinite dimensional Hilbert space $\Hh$. The cases with $d$ even and $\Hh$ being finite dimensional are dealt with in a similar manner. The crucial hypothesis is that the zero set $\Zz(H)$ only contains a finite number $I$ of singular points $x^*_1,\ldots,x^*_I{\in\Mm^d}$. In an ideal semimetal in the sense of Definition~\ref{def-IdealSemimet}, for any such singular point $x^*_i$, one has the local gauge transformation $x\mapsto W_{i,x}$ such that \eqref{eq-WeylPoint} holds. In this case, the argument in Section~\ref{sec-Weyl} shows that corresponding zero mode eigenstates of the spectral localizer are given by a Gaussian, see the proof of Proposition~\ref{prop-WeylSpecLocKernel}. The reader interested in this case only can jump directly to Proposition~\ref{prop-Semiclassics} below for the IMS localization argument. However, in Theorems~\ref{theo-CalliasSemiclassics} and \ref{theo-CalliasSemiclassicsInfDim} we only supposed that there is at most a linear vanishing of the eigenvalues, and the existence of the local gauge transformation has to be guaranteed. This is the object of the following lemma and then the next Proposition~\ref{prop-SemiclassicsOneWell} shows that the zero mode eigenstates have at least Gaussian decay.

\begin{lemma}
\label{lemma:finite_dim}
For every singular point $x^*_i$ there exists a ball $B_\delta(x^*_i)$ and some $a_i > 0$ such that $P_{i,x} = \chi(|H_x| \leq a_i)$ has constant finite rank $N_i$. Moreover there is a family of unitary  operators $W_i=(W_{i,x})_{x\in B_\delta(x^*_i)}$  with $W_{i,x}: \Hh \to (P_{i,x}^\perp \Hh) \oplus \CM^{N_i}$  such that 
\begin{equation}
\label{eq:finite_splitting}
W_{i,x} H_x W_{i,x}^* 
\;=\; \begin{pmatrix}
 P^\perp_{i,x} H_{x} P^\perp_{i,x} & 0 \\ 0 & H^F_{i,x} 
\end{pmatrix}
\;,
\end{equation}
with $x\in B_\delta(x^*_i) \mapsto H_{i,x}^F$ a differentiable matrix-valued function satisfying $|H_{i,x}^F| \geq c_1 |x-x_i^*|$. 
\end{lemma}

\noindent{\bf Proof.}
The assumptions on $H$ imply in particular that the resolvent maps $x \in \RM^d \mapsto (H_x - z)^{-1}$ are norm-differentiable whenever $z$ is in the resolvent set of $H_x$. The existence of $a_i, \delta>0$ is standard for a continuous family of Fredholm operators and from the Riesz projection formula one concludes that $x\in B_\delta(x_i^*) \mapsto P_{i,x}=\chi(|H_x| \leq a_i)$ is also a differentiable family of finite rank projections. One can then choose $N_i$ differentiable vector fields which are pointwise an orthonormal basis of $\Ran(P_{i,x})$, {\it e.g.} by choosing any basis $(\phi_1,...,\phi_{N_i})$ of  $\Ran(P_{i,x_i^*})$ and pointwise applying Gram-Schmidt orthogonalization to $(P_{i,x}\phi_1,...,P_{i,x}\phi_{N_i})$. This induces a family of isometries $I_{x}: \Ran(P_{i,x}) \to \CM^{N_i}$ for which $x\mapsto I_x H_x I_x^*$ is a differentiable matrix. Since $I_x I_x^*=\one_{N_i}$ and $I_x^*I_x=P_{i,x}$ hold by construction the unitary $W_{i,x} = I_x P_{i,x}+P^\perp_{i,x}$ then has the desired properties. In particular, $|H^F_{i,x}| \geq c_i |x-x_i^*|$ holds since the assumption  implies $|P_{i,x} H_x P_{i,x}| \geq c_i |x-x_i^*|$.
\hfill $\Box$

\vspace{.2cm}

By assumption, there exists a $a_i>0$ be such that $|P^\perp_{i,x}H_x P^\perp_{i,x}|\geq a_i\one$ for all $x\in\partial B_\delta(x^*_i)$ which assures uniform invertibility on the range of $P^\perp_i$. The finite-dimensional component can be decomposed further as
$$
H^F_{i,x}
\;=\;
H^Q_{i,x}\;+\;H^R_{i,x}
\;,
$$
with a linear term (leading to a square of the associated spectral localizer that has a harmonic quadratic potential, hence the letter $Q$) given by
$$
H^Q_{i,x}
\;=\; S_i \cdot (x-x^*_i) \;=\;
\sum_{j=1}^d S_{i,j} e_i \cdot (x-x^*_i)
\;,
$$
with the linearization matrix $S_i=(S_{i,1},\ldots,S_{i,d})$ given by $S_{i,j}=(\partial_j H^F)_{x^*_i}|_{\Ker(H^F_{x^*_i})}$  and a remainder $H^R_{i,x}$ satisfying $H^R_{i,x}=\Oo((x-x_i^*)^2)$. One sees that $|H^F_{i,x}| \geq c_i |x-x_i^*|$ is equivalent to the property that $\Ker(y\cdot S_i)=\{0\}$ for all $y\not=0$ which in turn is via a compactness argument on the unit sphere $\SM^{d-1}$ equivalent to $(y\cdot S_i)^2 > \tilde{c}_i |y|^2$ for some $\tilde{c}_i>0$. 


\vspace{.2cm}

The linear Hamiltonian $H^Q_{i,x}$ is now  viewed as a $N_i\times N_i$ matrix-valued function on all $\RM^d$. Then let $L^Q_{\kappa,i}$ be the associated spectral localizer. For the operator $H^Q_{i}$ one can apply Theorem~\ref{theo-CalliasFinite}, namely
$$
\Sig\big(J|_{\Ker(L^Q_{\kappa,i})}\big)
\;=\;
c^*_i
\;,
$$
where $c^*_i=\Ch_{d-1}(H^Q_i|H^Q_i|^{-1},\partial B_\delta(x^*_i))$ is the topological charge of $x^*_i$.
By homogeneity $\mult^*_i=\dim(\Ker(L^Q_{\kappa,i}))$ does not depend on $\kappa$ and one can choose an orthonormal frame  ${\Phi}_{\kappa,i}=(\phi_{\kappa,i,1},\ldots,\phi_{\kappa,i,\mult^*_i})\in L^2(\RM^d,\CM^{d'}\otimes\Hh)^{\mult^*_i}$ for $\Ker(L^Q_{\kappa,i})$ in the form $\Phi_{\kappa,i}(x)=\kappa^{-d/4} \Phi_{1,i}(x_i^* +(x-x_i^*)/\sqrt{\kappa})$. 
One then has $\mult^*_i\geq |c^*_i|$ with $\mult^*_i-c^*_i$ an even number and 
\begin{equation}
\label{eq-SigLocal}
c^*_i\;=\;\Sig({\Phi}_{\kappa,i}^*J{\Phi}_{\kappa,i})
\;.
\end{equation}
Since $H^Q_{i,x}$ is linear in $x$, one can see from \eqref{eq-LSquare2} that $(L^Q_{\kappa,i})^2$ is a bounded perturbation of $-\kappa^2\Delta + (S_i \cdot (X-x_i^*))^2$ which has a quadratically growing potential due to the assumption of linear growth. This then readily implies that $L_{\kappa,i}^Q$ has a compact resolvent. All nonzero eigenvalues of $L_{\kappa,i}^Q$ are also proportional to $\sqrt{\kappa}$ due to scaling. Let us recollect these findings in a preparatory result:

\begin{proposition}
\label{prop-SemiclassicsOneWell}
There is a constant $c_i>0$ such that the only spectrum of $L^Q_{\kappa,i}$ in $[-c_i\sqrt{\kappa},c_i\sqrt{\kappa}]$ is given by the kernel of dimension $\mult^*_i$ which satisfies $\mult^*_i\leq N_i$. Moreover, the eigenvectors decay like Gaussians: $\|e^{\mu \kappa^{-1} (X-x_i^*)^2} {\phi}_{\kappa,i,j}\|\leq A$ with $A,\mu>0$ independent of $\kappa$ and $j=1,\ldots,\mult^*_i$.
\end{proposition}

\noindent{\bf Proof.}
After the above remarks it only remains to prove the exponential decay of the eigenfunctions for $\kappa=1$ and shift the coordinate origin to $x_i^*=0$. Let us first derive an exponential decay for the resolvent by the Combes-Thomas method, closely following the presentations of \cite{His,Shu}. For simplicity let us write here $L=L_{1,H_i^Q}$ and define the analytic family 
$$
L(\lambda)\,: \, 
\lambda \in B_\delta(0) \subset \CM^d 
\;\mapsto\; 
e^{-\imath \lambda X^2} (L) e^{\imath \lambda X^2}
\; =\; 
L - 2 \imath \lambda \gamma\cdot X
\;,
$$
where we take the closures of these operators for $\delta$ small enough. This is well-defined since
the graph norm of $L$ is equivalent to $\sqrt{\|D \psi\|^2 + \|H^Q_i\psi\|^2+\|\psi\|^2} \simeq \sqrt{\|D \psi\|^2 + \|X\psi\|^2 +\|\psi\|^2}$ and hence $L-L(\lambda)$ is bounded relative to $L$ with $\mathrm{Dom}(L(\lambda))=\mathrm{Dom}(L)$ for $|\lambda|$ small enough.
 
\vspace{.1cm}

Let $E\in \CM$ be in the resolvent set of $L$ and moreover assume $\mathrm{dist}(E,\sigma(L))>c_1 > 0$ for some constant $c_1$. The existence of the relative bound implies that $\gamma\cdot X (L-E)^{-1}$ defines a bounded operator which satisfies the resolvent identities and hence there exists some $c_2 > 0$ such that $\|2\imath \gamma\cdot X (L-E)^{-1}\| < c_2 $ uniformly for all $E$ as above, see {\sl e.g.} \cite[Proposition 6.1.5]{dO}. One then has $L(\lambda)-E = (\one - 2\imath \lambda \gamma\cdot X (L-E)^{-1})(L-E)$ and both factors are invertible for $|\lambda| < \frac{1}{c_2}$, implying that one can arrange for $\|(L(\lambda)-E)^{-1}\| \leq c_3$
uniformly for all $E$ as specified above and small enough $|\lambda|$.

\vspace{.1cm}

For $\chi^r_x$ the multiplication operator for the characteristic function of a cube $\Lambda^r_x$ of sides $r$ centered in $r x\in \RM^d$ one has 
$$
\chi^r_x (L-E)^{-1} \chi^r_0 
\;=\; 
\chi^r_x e^{-\imath \lambda X^2} (L(\lambda)-E)^{-1} e^{\imath \lambda X^2} \chi^r_0
\;,
$$ 
which is obvious for real $\lambda$ and can then be continued analytically since the indicator functions provide a cutoff for the exponentials. Substituting purely complex values $\lambda =-\imath 2 c_3$ with $c_3>0$ small enough to be in the domain of analyticity of the resolvents this implies
$$
\|\chi^r_x (L-E)^{-1} \chi^r_0\| 
\;\leq\; 
c_4(r) \|\chi^r_x e^{-\imath c_3 r^2 x^2} (L(\lambda)-E)^{-1} \chi^\delta_0\| 
\;\leq\; 
c_5(r)  e^{-c_3 (rx)^2}
\;.
$$
In particular, due to the compact resolvent the constants can be chosen uniformly on a small circle around the isolated eigenvalue $0 \in \RM$, implying through the Riesz projection formula $\|\chi_x \chi(L = 0) \chi_0\| \leq c_5(r) e^{-(rx)^2}$ (this and the following holds for any other isolated eigenvalue $E_0$). Note that $\chi(L = 0)$ has finite rank $\mult^*_i$ and thus one can now fix some $r$ large enough such that $\chi(L = 0) \chi^r_0$ also has rank $\mult^*_i$.  Hence we can find a basis $\psi_1,...,\psi_{\mult^*_i}$ of the eigenspace in the form $\psi_j=\chi(L = 0) \chi_0^r \varphi_j$ for some normalized $\varphi_1,...,\varphi_{\mult^*_i}$. Due to 
$$
\|\chi^r_x \psi_j\|
\;\leq \;
\|\chi^r_x \chi(L = 0) \chi^r_0\|
\;\leq \;
c_5(r) e^{-c_3 (rx)^2}
\;,
$$
the Gram-Schmidt procedure applied to $\psi_1,...,\psi_{\mult^*_i}$ results in the claimed orthonormal basis which has the desired decay rate. 
\hfill $\Box$

\vspace{.2cm}

In an ideal Weyl semimetal the local model is a Weyl Hamiltonian and then ${\Phi}_{\kappa,i}$ are exact Gaussian states described in the proof of Proposition~\ref{prop-WeylSpecLocKernel}. In the following, it will now be shown that the kernels of these wells as found in Proposition~\ref{prop-SemiclassicsOneWell} produce the low-lying spectrum for the Schr\"odinger operator $(L_\kappa)^2$ smaller than $\kappa^{\frac{4}{3}}$, and that there are no further eigenvalues of smaller order than $\kappa$. Translated for $L_\kappa$ this becomes

\begin{proposition}
\label{prop-Semiclassics}
Suppose that the hypothesis of {{\rm Theorem~\ref{theo-SemiclassicsIntro}} hold if $\Mm^d=\TM^d$, and of {\rm Theorem~\ref{theo-CalliasSemiclassicsInfDim}} if $\Mm^d=\RM^d$}. Then there are constants $c$ and $C$ such that the spectrum of $L_\kappa$ in $[-c\kappa^{\frac{2}{3}},c\kappa^{\frac{2}{3}}]$ consists of eigenvalues $\nu_{\kappa,1},\ldots,\nu_{\kappa,\mult^*}$ {\rm (}enumerated with their multiplicity{\rm )} where $\mult^*=\sum_{i=1}^I \mult_i^*$. There is no further spectrum in $[-C\kappa^{\frac{1}{2}},C\kappa^{\frac{1}{2}}]$, namely 
$$
\sigma(L_\kappa)\,\cap\,[-C\kappa^{\frac{1}{2}},C\kappa^{\frac{1}{2}}] 
\;=\;
\{\nu_{\kappa,1},\ldots,\nu_{\kappa,\mult^*}\}
\;.
$$
\end{proposition}

\noindent {\bf Proof.} The argument is a modification of the IMS localization procedure as given by Simon \cite{Sim}. It is reproduced in \cite{CFKS}, see also Shubin \cite{Shu} for a treatment of the matrix-valued case.  Let us begin by building-up the IMS localization procedure. Choose a family of smooth functions $\chi_i^\delta:{\Mm^d}\to [0,1]$, depending on a parameter $\delta > 0$, as follows: Each $\chi_i^\delta$ is supported in the neighborhood $B_\delta(x^*_i)$ of a singular $x^*_i$ (where  we assume that $\delta$ is so small that $H_x$ is invertible on $B_\delta(x^*_i)\setminus\{x^*_i\}$) and such that assume that $\chi_i^\delta(x)=1$ for all $x\in B_{\frac{\delta}{2}}(x^*_i)$. By scaling we can and do make the choice uniformly in $i$ and $\delta$ such that
$$
|\partial_x\chi_i^\delta|\;\leq\; c\,\delta^{-1}
\;.
$$
Also let us introduce $\chi_0:{\Mm^d}\to[0,1]$ by
$$
(\chi^\delta_0)^2
\;=\;
1\,-\,\sum_{i=1}^{I}(\chi_i^\delta)^2
\;.
$$
All these functions are extended to the Hilbert space fibers and Clifford degrees of freedom by tensoring with the identity, and for sake of notational simplicity no new notation will be used. Due to  $[\chi_{i,x}^\delta,H_x]=0$ the ILS localization formula now reads
$$
(L_\kappa)^2
\;=\;
\chi_0^\delta
(L_\kappa)^2
\chi_0^\delta
\;+\;
\sum_{i=1}^{I}
\chi_i^\delta
(L_\kappa)^2
\chi_i^\delta
\;-\;
\kappa^2\sum_{i=0}^{I}
|\partial_x\chi_i^\delta|^2
\;.
$$
The operator $(L_\kappa)^2$ restricted to the support of  $\chi_0^\delta$ has a strictly positive diagonal potential bounded below by $c_1 \delta^2$ and an off-diagonal perturbation of order $\kappa$. Thus
$$
\chi_0^\delta
(L_\kappa)^2
\chi_0^\delta
\;\geq\; 
c_1 \delta^2 (\chi^\delta_0)^2
\,-\,c_2\,\kappa
\;.
$$
Combined with the bound on the derivative $\partial_x\chi_i^\delta$ one thus deduces
\begin{equation}
\label{eq-IntermedBound}
(L_\kappa)^2
\;\geq\;
c_1 \delta^2 (\chi^\delta_0)^2
\;-\;c_2\,\kappa
\;+\;
\sum_{i=1}^{I}
\chi_i^\delta
(L_\kappa)^2
\chi_i^\delta
\;-\;
c_3\kappa^2\delta^{-2} 
\;.
\end{equation}
In order to obtain a local toy model for $(L_\kappa)^2$ in a neighborhood of $x^*_i$, let us now replace \eqref{eq:finite_splitting}. 
Let us for each $i$ choose a unitary fibered operator {$W_i: L^2(\Mm^d; \Hh) \to \int_{\Mm^d}^\oplus dx \,\Hh_x $} with the fibers of $W_i=\int^\oplus dx\,W_{i,x}$ chosen as in Lemma~\ref{lemma:finite_dim} inside $B_\delta(x_i^*)$ and {continued outside arbitrarily but with uniformly bounded derivatives}.

\vspace{.1cm}

In the following we only use $W$ on the range of the cutoff function $\chi_i^\delta$ such that the block-decomposition \eqref{eq:finite_splitting} can simply be written as a direct sum, i.e. $\chi_i^\delta W_i H W_i^* \chi_i^\delta  = \chi_i^\delta (H^0_i \oplus H^F_i)\chi_i^\delta$ (which must be understood as a direct sum of two fields of Hilbert spaces since the decomposition depends on $x$). {We assume in the following that for each $i$ the term $x\in B_{\delta}(x_i^*)\mapsto H_i^0$ is the restriction of a uniformly invertible differentiable potential $x\in \RM^d\mapsto \tilde{H}^0_{i,x}$ on $\RM^d$ chosen independent of small enough $\delta>0$ (in the case $\Mm^d=\TM^d$ this involves an identification of balls in $\Mm^d$ with their isometric images in $\RM^d$). Such potentials can always be constructed by blowing up a large enough sphere via some differentiable bijection $f_i: B_{r}(x_i^*)\to \RM^d$, which has uniformly bounded derivative and fixes $B_{r/2}(x_i^*)$, and then setting $\tilde{H}^0_{i,x} = H^0_{i,f(x)}$.}

\vspace{.1cm}

All remaining {$2\times 2$ matrices are now w.r.t.} the grading induced by $J$, i.e. they result from the doubling employed in the definition \eqref{eq-SpecLocRedefine} of $L_\kappa$.  Then
\begin{align*}
  & \chi_i^\delta 
(L_\kappa)^2
\chi_i^\delta 
\;=\;
\chi_i^\delta W_i^*
[W_iL_\kappa
W_i^*]^2W_i
\chi_i^\delta
\\
&
\;=\;
\chi_i^\delta W_i^*
\left[
\begin{pmatrix}
0 & \kappa W_i {D} W_i^* 
\\
\kappa W_i {D} W_i^* & 0
\end{pmatrix}
\;+\;
\begin{pmatrix}
0& -\imath W_i{H}W_i^* 
\\
\imath W_i{H}W_i^* & 0
\end{pmatrix}
\right]^2 W_i
\chi_i^\delta
\\
&
\;=\;
\chi_i^\delta W_i^*
\left[
\begin{pmatrix}
0 & \kappa {D}   
\\
\kappa {D} & 0
\end{pmatrix}
\;+\;
\begin{pmatrix}
0& -\imath W_i{H}W_i^* 
\\
\imath W_i{H}W_i^* & 0
\end{pmatrix}
\;+\;
\begin{pmatrix}
0 & \kappa W_i [{D}, W_i^*]  
\\
\kappa W_i [{D}, W_i^*] & 0
\end{pmatrix}
\right]^2W_i
\chi_i^\delta
\\
&
\;=\;
\chi_i^\delta W_i^*
\left[
L^{0}_{\kappa,i}\oplus L^{Q}_{\kappa,i}
\;+\;
\begin{pmatrix}
0 & \kappa W_i [{D}, W_i^*] -\imath\, 0\oplus{H}^R_{i}
\\
\kappa W_i [{D}, W_i^*] +\imath\, 0\oplus{H}^R_{i} & 0
\end{pmatrix}
\right]^2W_i
\chi_i^\delta
\;,
\end{align*}
{where the cutoff imposed by $\chi_i^\delta$ allows to insert spectral localizers for potentials over $\RM^d$ (with slight abuse of notation) given by
$$
L^{0}_{\kappa,i}
\;=\;
\begin{pmatrix}
0 & \kappa {D} -\imath {\tilde{H}}_{i}^0
\\
\kappa {D}+\imath {\tilde{H}}_{i}^0 & 0
\end{pmatrix}
\;,
\qquad
L^{Q}_{\kappa,i}
\;=\;
\begin{pmatrix}
0 & \kappa {D}  -\imath {H}_{i}^Q
\\
\kappa {D}  +\imath {H}_{i}^Q & 0
\end{pmatrix}
\;.
$$}
Next recall that, for two selfadjoint operators $A$ and $B$ (here $A$ is unbounded and $B$ is bounded), one has has the operator Cauchy-Schwarz inequality $\{A,B\}\leq A^2+B^2$ so that
$$
(A+B)^2
\;=\;
A^2\,+\,B^2\,+\,\{\delta^{\frac{\eta}{2}} A,\delta^{-\frac{\eta}{2}}B\}
\;\geq\;
(1-\delta^{\eta})A^2\,+\,(1-\delta^{-\eta})B^2
\;,
$$
for $\eta>0$ to be chosen later (such that the second negative term becomes small due to a $\delta$-dependence of $B$). Therefore
\begin{align*}
\chi_i^\delta 
(L_\kappa)^2
\chi_i^\delta 
\;\geq\;&
(1-\delta^\eta)\,\chi_i^\delta W_i^* (L^{0}_{\kappa,i})^2 \oplus  (L^{Q}_{\kappa,i})^2 W_i\chi_i^\delta 
\\
&\;\,+\,
(1-\delta^{-\eta})\,\chi_i^\delta W_i^*
\begin{pmatrix}
0 & \kappa W_i [{D}, W_i^*] -\imath\, 0\oplus{H}^R_{i}
\\
\kappa W_i [{D}, W_i^*] +\imath\, 0\oplus{H}^R_{i} & 0
\end{pmatrix}
W_i
\chi_i^\delta
\;.
\end{align*}
The contribution from the gapped bands, $(L^{0}_{\kappa,i})^2$, has a mass gap that is uniformly (in $\kappa$) bounded from below so that
$$
\chi_i^\delta W_i^*(L^{0}_{\kappa,i})^2 W_i \chi_i^\delta 
\;\geq\;
c_4 (\chi_i^\delta)^2
\;.
$$
As to the second summand let us note that $[{D},W_i^*]$ is bounded in operator norm because $W_i$ is smooth. As ${H}^R_{i}$ is of the order $\delta^2$ on the support of $\chi_i^\delta$, one has
$$
\chi_i^\delta W_i^*
\begin{pmatrix}
0 & \kappa W_i [{D}, W_i^*] -\imath\, 0\oplus{H}^R_{i}
\\
\kappa W_i [{D}, W_i^*] +\imath\, 0\oplus{H}^R_{i} & 0
\end{pmatrix}^2 W_i
\chi_i^\delta
\;=\;
\Oo(\kappa^2,\kappa\delta^2 ,\delta^4)
\;.
$$
The main contribution therefore comes from $L^{Q}_{\kappa,i}$ which was already analyzed in Proposition~\ref{prop-SemiclassicsOneWell} and shown to have a kernel spanned by orthonormal vectors $\phi_{\kappa,i,1},\ldots,\phi_{\kappa,i,\mult_i^*} \in L^2(\RM^d \otimes \CM^{N_i})$ which decay exponentially. By setting $\Phi_{\kappa,i}=(\phi_{\kappa,i,1},\ldots,\phi_{\kappa,i,\mult_i^*})$ and
\begin{equation}
	\label{eq-ApproxKernelVec}
	{\Psi}_{\kappa,\delta,i}\;=\;
	\chi^\delta_i\,W_i^*(0\oplus {\Phi}_{\kappa,i})
	\;\in\;L^2({B_\delta(x_i^*)},\CM^{d'}\otimes\Hh)^{\mult^*_i}
	\;,
\end{equation}
they lead to approximate eigenvectors ${\Psi}_{\kappa,\delta,i}=(\psi_{\kappa,\delta,i,1},\ldots,\psi_{\kappa,\delta,i,\mult_i^*})$ {which are considered to be elements of $L^2(\Mm^d,\CM^{d'}\otimes\Hh)^{\mult^*_i}$ via an obvious identification}.
They are almost orthonormal as long as $\delta\gg \kappa^{\frac{1}{2}}$.
Since all non-vanishing eigenvalues are larger than $C^*\kappa$ one has $(L^{Q}_{\kappa,i})^2 + C_i^*\kappa \sum_{j=1}^{\mult_i}\psi_{\kappa,\delta,i,j}\psi_{\kappa,\delta,i,j}^* \geq C^*\kappa \one$
and thus by monotonicity
$$
\chi_i^\delta W_i^*(L^{Q}_{\kappa,i})^2W_i\chi_i^\delta + F_{\kappa,\delta,i}
\;\geq\;
C_i^*\kappa 
(\chi_i^\delta)^2
\;
\;.
$$
for the positive operator ${F}_{\kappa,\delta,i} = C^*_i \kappa \sum_{j=1}^{\mult_i}\psi_{\kappa,\delta,i,j}\psi_{\kappa,\delta,i,j}^*$ of rank $\mult^*_i$. Collecting the estimates, one thus has 
$$
\chi_i^\delta 
(L_\kappa)^2
\chi_i^\delta 
\;\geq\;
C_i^*\kappa (1-\delta^{\eta})
(\chi_i^\delta)^2
\;-\;
{F}_{\kappa,\delta,i}
\;+\;
(1-\delta^{-\eta})\,
\Oo(\kappa^2,\kappa\delta^2 ,\delta^4)
\;.
$$
Setting $F_{\kappa,\delta}=\sum_{i=1}^{I}F_{\kappa,\delta,i} $ which is of rank $\mult^*=\sum_{i=1}^I\mult^*_i$ and $C^*=\min\{C^*_1,\ldots,C^*_{I}\}$ one replaces into \eqref{eq-IntermedBound} to conclude
$$
(L_\kappa)^2
\;\geq\;
(c_1 \delta^2-c_2\kappa) (\chi^\delta_0)^2
\;+\;
C^*\kappa (1-(\chi_0^\delta)^2)(1-\delta^\eta)
\;-\;
C^* \kappa F_{\kappa,\delta}
\;+\;
(1-\delta^{-\eta})\,\Oo(\kappa^2,\kappa\delta^2 ,\delta^4)
\;-\;
c_3\kappa^2\delta^{-2} 
\;.
$$
Now the size of the balls is chosen to be $\delta=\kappa^\alpha$ with $2\alpha<1$. Then for $\eta\in(0,1)$
\begin{equation}
\label{eq:Lkappasq_lowerbound}
(L_\kappa)^2
\;\geq
\;
\,C^*\kappa \one
\;-\;
C^* \kappa F_{\kappa,\delta}
\;-\;
c_5 \,\kappa^{(4-\eta)\alpha}\,
\;-\;
c_3 \,\kappa^{2(1-\alpha)}
\;.
\end{equation}
Choosing any $\alpha=\frac{2}{6-\eta}$  is optimal as the both last terms are then of the same order. Thus $(L_\kappa)^2$ is bounded below by $\frac{1}{4}C^*\kappa \one$, up to a perturbation $F$ of finite rank $\mult^*$, which may lead to at most $\mult^*$ eigenvalues below $\frac{1}{4}C^*\kappa \one$.

\vspace{.2cm}

To exhibit that $(L_\kappa)^2$ also has at least $\mult^*$ eigenvalues in $[0,c\kappa^{\frac{4}{3}}]$ we use the trial functions $\Psi_{\kappa,\delta,i}\in L^2({\Mm^d},\CM^{d'}\otimes\Hh)^{\mult^*_i}$ (again closely following \cite{Sim}).
Similar as above (with $\eta=0$), the operator Cauchy-Schwarz inequality implies that
$$
\chi_i^\delta 
(L_\kappa)^2
\chi_i^\delta 
\;\leq\;
2\,\chi_i^\delta\,W_i^*\,
(L^{0}_{\kappa,i}\oplus L^{Q}_{\kappa,i})^2
\,W_i\,\chi_i^\delta
\;+\;
2\,\Oo(\kappa^2,\kappa\delta^2 ,\delta^4)
\;.
$$
Hence, using as in the proof of the ILS localization formula $\chi \partial^2\chi=\chi^2\partial^2+\partial^2\chi^2+2(\partial\chi)^2$,
\begin{align*}
\langle {\Psi}_{\kappa,\delta,i}| 
(L_\kappa)^2
|{\Psi}_{\kappa,\delta,i} \rangle
&
\;=\;
\langle {\Phi}_{\kappa,i}|\,\chi_i^\delta\,
(L^{Q}_{\kappa,i})^2
\,\chi_i^\delta\,|{\Phi}_{\kappa,i}\rangle
\\
&
\;=\;
0\;+\;
\Oo(\kappa^2\delta^{-2})
\;+\;
\Oo(\kappa^2,\kappa\delta^2 ,\delta^4)
\;,
\end{align*}
which is understood as a $q^\mult_i\times q^\mult_i$ matrix.
Choosing again $\delta=\kappa^\alpha$ now with $\alpha=\frac{1}{3}$, it follows that
$\langle {\Psi}_{\kappa,i}| (L_\kappa)^2 |{\Psi}_{\kappa,i}\rangle =\Oo(\kappa^{\frac{4}{3}})$ for all $i=1,\ldots,I$. Since the trial eigenfunctions are almost orthonormal, the Rayleigh-Ritz principle implies that there are thus at least $\mult^*=\sum_{i=1}^I\mult^*_i$ eigenvalues of $(L_\kappa)^2$ that are smaller than $\Oo(\kappa^{\frac{4}{3}})$. This implies the claim.
\hfill $\Box$

\vspace{.2cm}

\noindent {{\bf Proof of Theorem~\ref{theo-SemiclassicsIntro}.} The claim on the spectrum of $L_\kappa$ is precisely Proposition~\ref{prop-Semiclassics} in the case $\Mm^d=\TM^d$.}
\hfill $\Box$

\vspace{.2cm}

\noindent {{\bf Proof of Theorem~\ref{theo-CalliasSemiclassics} as well as Theorem~\ref{theo-CalliasSemiclassicsInfDim}.} 
The claim on the spectrum of $L_\kappa$ in Theorem~\ref{theo-CalliasSemiclassics} is precisely Proposition~\ref{prop-Semiclassics} for $\Mm^d=\RM^d$. For the proof of the identity \eqref{eq-RobSal}} let $\Upsilon_\kappa=(\Upsilon_{\kappa,1},\ldots,\Upsilon_{\kappa,I}):\CM^m\to L^2(\RM^d,\CM^{d'}\otimes\Hh)^m$ be a partial isometry onto the normalized eigenstates $\Upsilon_{\kappa,i}\in L^2(\RM^d,\CM^{d'}\otimes\Hh)^{\mult^*_i}$ corresponding to the low-lying spectrum of $L_\kappa$ as identified in the proof of Proposition~\ref{prop-Semiclassics}, namely
$$
\Upsilon_\kappa^*L_\kappa \Upsilon_\kappa
\;=\;
\diag(\nu_{\kappa,1},\ldots,\nu_{\kappa,m})
\;.
$$
As all low-lying eigenvalues are included (see Proposition~\ref{prop-Semiclassics}), the identity $JL_\kappa J=-L_\kappa$ implies that $\{\nu_{\kappa,1},\ldots,\nu_{\kappa,m}\}$ is invariant under reflection and that the set $\Upsilon_{\kappa,1},\ldots,\Upsilon_{\kappa,I}$ of eigenvectors is $J$-invariant, so the projection $P_\kappa = \Upsilon_{\kappa}\Upsilon_{\kappa}^*$ commutes with $J$. As in the proof of Proposition~\ref{prop-Semiclassics} we have a semiclassical approximation to $P_\kappa$ given by the projection $\mathscr{P}_{\kappa,\delta}$ to the span of the approximate eigenvectors $\Psi_{\kappa,\delta,i}$. Indeed, fixing $\delta=\kappa^{\frac{1}{3}}$ and using that the approximate eigenvectors are almost orthonormal, one has $\mathscr{P}_{\kappa,\delta}=F_{\kappa,\delta} + \Oo(\kappa^\gamma)$ for arbitrarily large $\gamma>0$ and with $F_{\kappa,\delta}$ as in the proof of Proposition~\ref{prop-Semiclassics}.  In this situation where one has the projection to an approximate eigenspace for eigenvalues that are separated from the rest of the spectrum by a distance $\Oo(\kappa^{\frac{1}{2}})$, one can apply the estimate \cite[Proposition 2.5]{HS84} to show
$$
\lVert P_\kappa \mathscr{P}_{\kappa,\delta} \,-\, \mathscr{P}_{\kappa,\delta}\rVert 
\;\leq \;
\Oo( \kappa^{\beta_1})
$$
for some exponent $0 < \beta_1 < \frac{1}{2}$. On the other hand, as $F_\kappa=\mathscr{P}_{\kappa,\delta} + \Oo(\kappa^2)$ it follows from \eqref{eq:Lkappasq_lowerbound} that
$$
\langle \phi| L^2_\kappa \phi\rangle \,+\, C^* \kappa \, \langle \phi| \mathscr{P}_{\kappa,\delta}\phi\rangle 
\;\geq \;
C^*\kappa \,+\, \Oo(\kappa^{\frac{4}{3}})
$$
for any unit vector $\phi$ in range of $P_\kappa$, hence  $\langle \phi| \mathscr{P}_{\kappa,\delta}\phi\rangle \geq 1 + \Oo(\kappa^{\frac{1}{3}})$ since $\langle \phi| L^2_\kappa \phi\rangle\leq \Oo(\kappa^{\frac{4}{3}})$ by definition. One concludes that also $\lVert \mathscr{P}_{\kappa,\delta} P_\kappa - P_\kappa \rVert \leq c \kappa^{\beta_2}$ for some $0 < \beta < \frac{1}{2}$. Finally, that implies 
$$
\lVert P_\kappa - \mathscr{P}_{\kappa,\delta}\rVert 
\;=\; 
\lVert (\mathscr{P}_{\kappa,\delta} P_\kappa - \mathscr{P}_{\kappa,\delta})+ (P_\kappa \mathscr{P}_{\kappa,\delta} -P_\kappa)\rVert^{\frac{1}{2}} 
\;\leq \;
\Oo(\kappa^{\frac{1}{2}\beta_1})\,+\,\Oo(\kappa^{\frac{1}{2}\beta_2})
\;.
$$
Since $\mathscr{P}_{\kappa,\delta}$ is also $J$-invariant, it follows that
$$
\Sig(\Upsilon_\kappa^*J\Upsilon_\kappa)
\;=\; 
\Tr(J P_\kappa)\;=\; \Tr(J\mathscr{P}_{\kappa,\delta})
\;=\;
\sum_{i=1}^I \Sig({\Phi}_{\kappa,i}^*J{\Phi}_{\kappa,i}) 
\;=\;
\sum_{i=1}^I 
c^*_i
\;
$$
due to \eqref{eq-SigLocal} and since all terms are integers up to errors that vanish for $\kappa \to 0$.
The span of $\Upsilon_\kappa$ differs from $\Ker(L_\kappa)$ at most by the finite-dimensional eigenspaces corresponding to non-zero eigenvalues $\nu$ of $L_\kappa$ which come in symmetric pairs $(\nu,-\nu)$ with eigenvectors also given by pairs  $(\phi,J\phi)$ of vectors. Each of those two-dimensional $J$-invariant subspaces has a vanishing $J$-signature. Therefore, while the exact kernel of $L_\kappa$ and its semiclassical approximation do not necessarily have equal dimensions, they always have the same spectral asymmetry w.r.t. $J$, thus \eqref{eq-SusyInd} implies
\begin{equation}
\label{eq-IndSig}
\Ind(D_{\kappa,H})
\;=\;
\Sig\big(J|_{\Ker(L_\kappa)}\big)
\;=\;
\Sig\big(\Upsilon_\kappa^*J\Upsilon_\kappa\big)
\;=\;
\sum_{i=1}^I 
c^*_i
\;.
\end{equation}
This was proved for $\kappa>0$ sufficiently small, but then $\kappa$ can be raised up to $1$ without harming the Fredholm property and thus changing the index. This completes the proof.
\hfill $\Box$

\section{Chern integrals}
\label{sec-ChernIntegrals}

The main objective of this section is to prove Proposition~\ref{prop-ChernDiff} and then, in the second part of the section, to provide the tools to prove the even dimensional equivalent. The proof of Proposition~\ref{prop-ChernDiff} requires the computation of the odd Chern number $Q_{\mbox{\rm\tiny eff}}(m)$. This effective flat band Hamiltonian is off-diagonal with unitary off-diagonal entries. According to \eqref{eq-PassageOddEven}, this lower entry is given by
\begin{equation}
\label{eq-UDef}
{U}_k(m)\;=\;({H}_k\,+\,\imath\,m\,{{Y}}_k)\,|{H}_k\,+\,\imath\,m\,{{Y}}_k|^{-1}
\;.
\end{equation}
By construction,  the matrix ${U}_k(m)$ is well-defined and unitary for $m\not=0$ sufficiently small. Let $m_c$ be the smallest positive number such that $H+\imath m_c {Y}$ is not invertible. Note that the gap of ${H}_k+\imath\,m\,{{Y}}_k)$ grows linearly in $m$ uniformly in $k$ for small $m$ (because ${{Y}}_k$ does not vanish at the Weyl points), therefore $m_c>0$. Thus $|H+\imath m{Y}|\geq c |m|$ for $m$ small. In this connection let us stress that the lower bound holds even though ${Y}$ is not invertible. Next, by a well-known elementary computation ({\it e.g.} \cite{CSB}), the odd Chern number of $\Ch_d(Q_{\mbox{\rm\tiny eff}}(m))$ can be expressed in terms of this unitary and is equal to
\begin{align*}
{\mathrm{Ch}}_{d}(U(m)) 
\;=\;
\frac{\imath(\imath \pi)^\frac{d-1}{2}}{ d!!} 
\sum_{\rho \in \Ss_d} (-1)^\rho\int_{\TM^d} \frac{dk}{(2\pi)^d} \;
\mathrm{Tr}\left ( 
 \prod_{j=1}^{d}   {U}_k(m)^\ast \, \partial_{k_{\rho(j)}}{U}_k(m)
\right ) \; .
\end{align*}

\vspace{.2cm}

\noindent {\bf Proof of Proposition~\ref{prop-ChernDiff}.} 
Recall that  $k_1^*,\ldots,k^*_{I}$ denote the Weyl points of the ideal Weyl semimetal $H$ and that $c^*_1,\ldots,c^*_{I}$ are their charges. {For sake of simplicity, let us assume that all Weyl points are simple so that the charges satisfy $|c_i^*|=1$ for all $i$.} Let now $\chi_i^\delta$ with $i=1,\ldots,I$ be smooth positive functions that are equal to $1$ on ${B}_{\frac{\delta}{2}}(k^*_i)$ and vanish outside the ball $B_{\delta}(k^*_i)$. Further let $\chi^\delta_0=1-\sum_{i=1}^{I}\chi_i^\delta$ complete the partition of unity of $\TM^d$. Then 
\begin{align*}
{\mathrm{Ch}}_{d}(U(m)) 
&
\;=\;
\frac{\imath(\imath \pi)^\frac{d-1}{2}}{ d!!} \;\sum_{i=0}^{I}
\sum_{\rho \in \Ss_d} (-1)^\rho\int_{\TM^d} \frac{dk}{(2\pi)^d}  \;\chi_i^\delta(k)\;
\mathrm{Tr}\left ( 
 \prod_{j=1}^{d}   {U}_k(m)^\ast \, \partial_{k_{\rho(j)}}{U}_k(m)
\right ) \; .
\end{align*}
Now in the summand corresponding to $i=0$, the limit $m\to 0$ can safely be taken because $\lim_{m\to 0}{U}_k(m)={U}_k(0)$ exists for $k$ being outside of the Weyl points. Therefore the corresponding contributions in $ {\mathrm{Ch}}_{d}(U(\pm m)) $ are equal, so that for $m>0$
\begin{align*}
{\mathrm{Ch}}_{d}& (U(m)) \,-\,{\mathrm{Ch}}_{d}(U(-m)) \, +\,o(m)
\\
&
\,=\, 
\sum_{i=1}^{I}
\,\sum_{\eta=\pm 1}\eta \, 
\frac{\imath(\imath \pi)^\frac{d-1}{2}}{ d!!}
\,\sum_{\rho \in \Ss_d} (-1)^\rho\int_{B_\delta(k^*_i)} \frac{dk}{(2\pi)^d} \;\chi_i^\delta(k)\,
\mathrm{Tr}\left ( 
 \prod_{j=1}^{d}   {U}_k(\eta m)^\ast \, \partial_{k_{\rho(j)}}{U}_k(\eta m)
\right )\,.
\end{align*}
Now for each $i$ (that is each Weyl point $k^*_i$), one can (locally on $B_\delta(k^*_i)$) use the real analytic basis change $W_k$ from \eqref{eq-WeylPoint}. Replacing the basis change leads to numerous terms, but since the derivatives of $\partial_{k_j} W_k$ are of order $O(\delta)$ the terms involving them can be neglected if one chooses $\delta=m^{\frac{1}{2}}$. Expanding \eqref{eq-WeylPoint}  terms containing ${H}^0_k$ or $H_k^R$ are also irrelevant in the limit due to their scaling in $k$ and thus $\delta$. Since only the contributions coming from the linear term $H^Q_k$ remain it is enough to evaluate the integral for $U$ replaced by 
$$
U^W(m)
\;=\;
(H^W+\imath m)(H^W-\imath m)^{-1}
\;
$$ 
for {$H^W=\sum_{j=1}^d\langle k-k^*|B_ie_j\rangle \Gamma_j$} with chirality {$c^*_i=\sgn(\det(B_i))$}. By shifting we may assume $k^*_i=0$. Let us now introduce the associated Chern integral
\begin{equation}
\label{eq-ChernIntWeyl}
{\mathrm{C}}_d(U^W(m))
\;=\;
\frac{\imath(\imath \pi)^\frac{d-1}{2}}{ d!!}
\,\sum_{\rho \in \Ss_d} (-1)^\rho\int_{\RM^d} \frac{dk}{(2\pi)^d} \,
\mathrm{Tr}\left ( 
 \prod_{j=1}^{d}   {U}^W_k( m)^\ast \, \partial_{k_{\rho(j)}}{U}^W_k(m)
\right )\,.
\end{equation}
A careful analysis of the above integral shows that in the limit $m\to 0$ all contributions outside the ball $B_\delta(0)$, still with $\delta=m^{\frac{1}{2}}$, tend to $0$ and so the reductions take the final form
$$
\lim_{m\to 0}\big({\mathrm{Ch}}_{d} (U(m)) \,-\,{\mathrm{Ch}}_{d}(U(-m))\big)
\;=\;
\lim_{m\to 0}
\;
\sum_{i=1}^{I}
\,\sum_{\eta=\pm 1}\eta \;
{\mathrm{C}}_d(U^W(m\eta {y^*_i}))
\;.
$$
In Proposition~\ref{prop-ChernIntWeyl} below it is shown that ${\mathrm{C}}_d(U^W(m))=\frac{1}{2}(-1)^{\frac{d-1}{2}}\sgn(m)c_i^*$.  Hence the contribution to the difference of Chern numbers  stemming from this simple Weyl point is {$\frac{1}{2}y^*_ic^*_i-(-\frac{1}{2}y^*_ic^*_i)=y^*_ic^*_i$}. Summing over all summands in \eqref{eq-WeylPoint} and then over all singular points, one obtains the claim.
\hfill $\Box$

\vspace{.2cm}

It thus remains to compute the Chern integral \eqref{eq-ChernIntWeyl}. {As in Section~\ref{sec-ReductionWeyl}, one can first reduce the computation to the case $H^W=\sum_{j=1}^d k_j b_j  \Gamma_j$. Let us then} begin by rewriting the unitary:
\begin{align*}
{U}^W_k(m)
&
\;=\;
\left(\sum_{j=1}^d {b}_j \Gamma_j k_j+\imath m\right)
\left|\sum_{j=1}^d {b}_j \Gamma_j k_j+\imath m\right|^{-1}
\\
&
\;=\;
\left(\sum_{j=1}^d {b}_j \Gamma_j k_j+\imath m\right)
\left(\sum_{j=1}^d {b}_j^2 k_j^2+m^2\right)^{-\frac{1}{2}}
\end{align*}
It is clear from this expression that the limit $\lim_{|k|\to\infty}{U}^W_k(m)$ does not exist, as the limit points depend on the direction. Hence $k\in\RM^d\mapsto {U}^W_k(m)$ cannot be compactified and does not contain topological information. Nevertheless, one can compute the associated differential form in \eqref{eq-ChernIntWeyl}. This differential form turns out to be integrable with a computable associated integral ${\mathrm{C}}^W_{d}(m)$.


\begin{proposition}
\label{prop-ChernIntWeyl}
For $m\not=0$, the Chern integral of the Weyl Hamiltonian $H^W_k=\sum_{j=1}^d{b}_jk_j\Gamma_j$ defined in \eqref{eq-ChernIntWeyl} is equal to
$$
{\mathrm{C}}^W_d(m)
\;=\;
\frac{1}{2}\,(-1)^{\frac{d-1}{2}}\,\sgn({b})\,\sgn(m)
\;,
$$
where $\sgn({b})=\prod^d_{j=1}\sgn({b}_j)$ is the chirality of the Weyl Hamiltonian.
\end{proposition}

\noindent {\bf Proof.} 
For sake of notational simplicity, let us suppress the index $W$ on ${U}^W_k(m)$. Replacing the above expression for ${U}_k(m)$, one can first of all make a change of variables ${b}_jk_j\mapsto k_j$. Then
\begin{align*}
{\mathrm{C}}_{d}(U(m)) 
\;=\; 
\frac{\imath(\imath \pi)^\frac{d-1}{2}}{ d!!}\; \sgn({b}) \;
\sum_{\rho \in \Ss_d} (-1)^\rho\int_{\RM^d} \frac{dk}{(2\pi)^d} \;
\mathrm{Tr}\left ( \prod_{j=1}^d 
  {U}_k^\ast(m) \, \partial_{k_{\rho(j)}}{U}_k(m)
\right ) \; ,
\end{align*}
where now ${U}(m)$ is given by the formula above with ${b}=(1,\ldots,1)$. One finds
$$
\partial_{k_{j}}{U}_k(m)
\;=\; \frac{1}{\sqrt{k^2+m^2}} \Big(\Gamma_j - (\Gamma_0 \cdot k + \imath m) \frac{k_j}{k^2+m^2}\Big)
\;.
$$
%
A trick to simplify the trace is to note that the computation can be embedded into a representation of the odd-dimensional Clifford algebra $\CM_{d+2}$ with $d+2$ generators given by
$$
\sigma_{k}
\;=\;\begin{pmatrix}
0 & \Gamma_k \\ \Gamma_k & 0
\end{pmatrix}
\;, 
\qquad 
\sigma_{d+1}
\;=\;
\imath \begin{pmatrix} 0 & -\one \\ \one & 0 \end{pmatrix}
\;, 
\qquad 
\sigma_{d+2}
\;=\;
\begin{pmatrix} \one & 0 \\ 0 & -\one \end{pmatrix}
\;.
$$ 
Indeed, introducing the vectors $q=(k_1,...,k_d, m,0) \in \mathbb{R}^{d+2}$ and $q_j=e_j - k_j \frac{q}{|q|^2}$, one recognizes
$$
\sigma\cdot q 
\;=\; 
\lvert q\rvert 
\begin{pmatrix}
	0 & \hat{U}_k^*(m)\\ \hat{U}_k(m) & 0
\end{pmatrix}
\;, 
\qquad 
\sigma\cdot q_j 
\;=\; 
\lvert q\rvert 
\begin{pmatrix}
	0 & \partial_{k_{j}}\hat{U}_k^*(m)\\ \partial_{k_{j}}\hat{U}_k(m) & 0
\end{pmatrix}
\;,
$$ 
and hence using Lemma~\ref{lemma:cliffordtrace} below one can compute
\begin{align*}
\mathrm{Tr}\left ( 
{U}_k^\ast \prod_{j=1}^{d} \, \partial_{k_{\rho(j)}}{U}^{(j+1)*}_k
\right )
&
\;=\;
 \frac{1}{\lvert q\rvert^{d+1}}\,
\Tr\Big((\sigma\cdot q)\big((\sigma\cdot q_{\rho(1)})\cdots (\sigma \cdot q_{\rho(d)})\big) \frac{1}{2}\,(\one+\sigma_{d+2})\Big) \\
&
\;=\; 
2^{\frac{d-1}{2}}\imath^{\frac{d+1}{2}} \frac{1}{\lvert q\rvert^{d+1}}\,
\det\big(q,q_{\rho(1)},\ldots,q_{\rho(d)},e_{d+2}\big)\\
&
\;=\;
2^{\frac{d-1}{2}}
\,\imath^{\frac{d+1}{2}} 
\,\frac{1}{\lvert q\rvert^{d+1}}
\;
\det\big(q,e_{\rho(1)},\ldots,e_{\rho(d)},e_{d+2}\big)\\
&
\;=\;
-\,2^{\frac{d-1}{2}}
\,\imath^{\frac{d+1}{2}}\, \frac{1}{\lvert q\rvert^{d+1}}\,(-1)^\rho\, m
\;.
\end{align*} 
Note, in particular, that after taking the trace one ends up with an absolutely convergent integrand. As $S_d$ has $d!$ elements, one thus has
$$
{\mathrm{C}}_{d}(U(m)) 
\;=\; 
-\frac{\imath(\imath \pi)^\frac{d-1}{2}}{ d!!}\; \sgn(s) \;
d!\; 2^{\frac{d-1}{2}}\imath^{\frac{d+1}{2}}
\int_{\RM^d} \frac{dk}{(2\pi)^d} \;\frac{m}{(k^2 + m^2)^{\frac{d+1}{2}}}
\;.
$$
In the integral one can change variables $k\mapsto \frac{k}{m}$, producing a factor $\sgn(m)^d=\sgn(m)$:
$$
{\mathrm{C}}_{d}(U(m)) 
\;=\; 
(-1)^{\frac{d-1}{2}}\frac{\pi^\frac{d-1}{2}}{ d!!}\; \sgn(s) \;
d!\,2^{\frac{d-1}{2}}\,\sgn(m)\;\frac{1}{(2\pi)^d}
\int_{\RM^d} dk \;\frac{1}{(k^2 + 1)^{\frac{d+1}{2}}}
\;.
$$
Then the integral can be evaluated in polar coordinates:
$$
\int_{\RM^d} dk \;\frac{1}{(k^2 + 1)^{\frac{d+1}{2}}}
\;=\;
\mbox{Vol}(\SM^{d-1})
\int^\infty_0 dr \;\frac{r^{d-1}}{(r^2 + 1)^{\frac{d+1}{2}}}
\;=\;
\mbox{Vol}(\SM^{d-1})
\;\frac{\sqrt{\pi}\;\Gamma(\frac{d}{2})}{2\;\Gamma(\frac{d+1}{2})}
\;,
$$
where here $\Gamma$ now denotes the $\Gamma$-function, so that $\Gamma(\frac{d+1}{2})=\frac{d-1}{2}!$. Furthermore, 
$$
\mbox{Vol}(\SM^{d-1})\;=\;\frac{2\pi^{\frac{d}{2}}}{\Gamma(\tfrac{d}{2})}
\;,
\qquad
d!\;=\;2^{\frac{d-1}{2}} d!! \tfrac{d-1}{2}!
\;.
$$ 
Carefully tracking all factors, one can conclude the proof.
\hfill $\Box$
%
%

\begin{lemma}
\label{lemma:cliffordtrace}
Let $\sigma_1,...,\sigma_{n}$ with $n$ odd be {a left-handed} irreducible representation of the generators of $\CM_n$.
Then for column vectors $q_1,...,q_n \in \mathbb{R}^n$
$$
\Tr\left(\prod_{k=1}^{n} \sigma\cdot q_k\right)
\;=\;
2^{\frac{n-1}{2}}\imath^{\frac{n-1}{2}}\det(q_1,...,q_{n})
\;.
$$
\end{lemma} 

\noindent {\bf Proof.}
First of all,  $\sigma_1\cdots\sigma_d=\imath^{\frac{n-1}{2}}\one$ where the $\one$ is the identity on the representation space of dimension $2^{\frac{n-1}{2}}$. Furthermore, the trace of a product of an odd number $m<n$ of matrices $\sigma_j$ always vanishes. From these facts the claim follows.
\hfill $\Box$

\vspace{.2cm}

Let us finally come to the case of even dimension $d$. Thus the Hamiltonian is supposed to satisfy $H=-\Gamma_0 H\Gamma_0$. The self-adjoint operator $H+m\Gamma_0$ is invertible because $(H+m\Gamma_0)^2=H^2+m^2$ and therefore there is an associated spectral projection on the negative spectrum:
$$
P(m)
\;=\;
\frac{1}{2}
\big(
\one\,-\,(H+m\Gamma_0)|H+m\Gamma_0|^{-1}\big)
\;.
$$
When computing the difference of the Chern numbers $\Ch_d(P(m))-\Ch_d(P(-m))$ one is hence naturally led to study the contributions of the Dirac Hamiltonian $H^D_k=\sum_{j=1}^d{b}_jk_j\Gamma_j$. Its projection are explicitly given by
\begin{align*}
{P}^D_k(m)
&
\;=\;
\frac{1}{2}
\left(
\one\,-\,
\Big(\sum_{j=1}^d {b}_j \Gamma_j k_j+m\Gamma_0\Big)
\Big(\sum_{j=1}^d {b}_j^2 k_j^2+m^2\Big)^{-\frac{1}{2}}
\right)
\;.
\end{align*}
Like the unitary phase of the Weyl operator, this family of projections does not extend to the one-point compactification since the limits as $|k|\to\infty$ are direction-dependent, hence it carries no well-defined Chern number. Nevertheless, let us consider its $d$th Chern integral given by
\begin{align*}
{\mathrm{C}}_{d}(P^D(m)) 
\;=\; 
\frac{(2\imath \pi)^\frac{d}{2}}{ \frac{d}{2}!} 
\sum_{\rho \in \Ss_d} (-1)^\rho\int_{\RM^d} \frac{dk}{(2\pi)^d} \;
\mathrm{Tr}\left ( 
 {P}_k(m) \prod_{j=1}^{d}  \partial_{k_{\rho(j)}}{P}_k(m)
\right ) \; .
\end{align*}
%

\begin{proposition}
\label{prop-ChernIntDirac}
Let $d$ be even. For $m\not=0$, the Chern integral of the Dirac Hamiltonian $H^D_k=\sum_{j=1}^d{b}_jk_j\Gamma_j$ defined in \eqref{eq-ChernIntWeyl} is equal to
$$
{\mathrm{C}}_{d}(P^D(m)) 
\;=\;
\frac{1}{2}\,(-1)^{\frac{d}{2}+1}\, \sgn({b})\,\sgn(m)
\;,
$$
where $\sgn({b})=\prod^d_{j=1}\sgn({b}_j)$ is a sign factor comparing $H^D$ with a standard Dirac Hamiltonian $H_k=\sum_{j=1}^d k_j\Gamma_j$  {\rm (}{such that $(-\imath)^{d/2}\Gamma_1 \cdots \Gamma_d = \Gamma_0${\rm )}}.
\end{proposition}

\noindent {\bf Proof.} 
Let us suppress the upper index $D$ on ${P}^D_k(m)$.
Replacing the above expression for ${P}_k(m)$, one can first of all make a change of variables ${b}_jk_j\mapsto k_j$. Then
\begin{align*}
{\mathrm{C}}_{d}(P(m)) 
\;=\; 
\frac{(2\imath \pi)^\frac{d}{2}}{ \frac{d}{2}!} 
\; \sgn({b}) \;
\sum_{\rho \in \Ss_d} (-1)^\rho\int_{\RM^d} \frac{dk}{(2\pi)^d} \;
\mathrm{Tr}\left ({P}_k(m) \prod_{j=1}^d 
\partial_{k_{\rho(j)}}{P}_k(m)
\right ) \; ,
\end{align*}
where now ${P}(m)$ is given by the formula above with ${b}=(1,\ldots,1)$.
For the computation of the integrand, let us rewrite 
$$
{P}_k(m) 
\;=\; 
\frac{1}{2} \Big(\one - \frac{\sigma \cdot q}{\lvert q \rvert}\Big)
\;,
$$
where $q=(k_1,...,k_d,m)$ and $\sigma=(\Gamma_1,...,\Gamma_d, \Gamma_0)$ are generating the odd Clifford algebra $\CM_{d+1}$. Then 
$$
\partial_{k_j}{P}_k
\;=\; 
-\,\frac{1}{2}
\;
\frac{1}{\lvert q \rvert}\,
\Big(\sigma_j - k_j \frac{\sigma\cdot q}{q^2}\Big) 
\;=\; 
-\,\frac{1}{2}\;\frac{1}{\lvert q \rvert}\; \sigma\cdot q_j
\;,
$$
with $q_j=e_j - k_j \frac{q}{q^2}\in \mathbb{R}^{d+1}$. The scalar term of ${P}_k(m)$ only results in products of at most $d$ of the $\sigma$-matrices which thus have vanishing trace, therefore we compute
\begin{align*}
\mathrm{Tr}\left ( 
\Big({P}_k(m)-\frac{1}{2}\one\Big) 
\prod_{j=1}^{d}  \partial_{k_{\rho(j)}}{P}_k(m)
\right ) 
&
\;=\; 
-\,\frac{1}{2^{d+1}}\;\frac{1}{\lvert q \rvert^{d+1}} \;
\Tr\Big((\sigma \cdot q)(\sigma \cdot q_{\rho(1)})...(\sigma \cdot q_{\rho(d)})\Big) 
\\
&
\;=\; 
-\,\frac{2^{\frac{d}{2}}}{2^{d+1}}
\;
\frac{\imath^{\frac{d}{2}}}{\lvert q\rvert^{d+1}}\; \det(q,q_{\rho(1)},\ldots,q_{\rho(d)})
\\
&
\;=\;
-\,2^{-\frac{d}{2}-1}\;\frac{\imath^{\frac{d}{2}}}{\lvert q\rvert^{d+1}} \;\det(q,e_{\rho(1)},\ldots,e_{\rho(d)})
\\ 
&
\;=\;
-\,2^{-\frac{d}{2}-1}\,\frac{\imath^{\frac{d}{2}}}{\lvert q\rvert^{d+1}} \,m \,(-1)^\rho
\;,
\end{align*}
where Lemma~\ref{lemma:cliffordtrace} was used to evaluate the trace. Hence 
\begin{align*}
{\mathrm{C}}_{d}(P(m)) 
\;=\; 
-\,2^{-\frac{d}{2}-1}\;\imath^{\frac{d}{2}}\; 
\frac{(2\imath \pi)^{\frac{d}{2}}}	{\frac{d}{2}!} 
\; \sgn(s) \; \int_{\RM^d} \frac{\mathrm{d}k}{(2\pi)^d} \;d!\;
\frac{m}{(k^2+m^2)^{\frac{d+1}{2}}}\; ,
\end{align*}
such that scaling by $m$ and going to polar coordinates leads to
$$
{\mathrm{C}}_{d}(P(m)) 
\;=\; 
-2^{-\frac{d}{2}-1}\imath^{\frac{d}{2}} d!
\frac{(2\imath \pi)^\frac{d}{2}}{ \frac{d}{2}!} \;
\sgn(s)\,\sgn(m) \;\frac{1}{(2\pi)^d}\;
\mbox{Vol}(\SM^{d-1})
\int^{\infty}_0dr\;\frac{r^{d-1}}{(r^2+1)^{\frac{d+1}{2}}}
\;.
$$
Finally using
$$
\int^{\infty}_0dr\;\frac{r^{d-1}}{(r^2+1)^{\frac{d+1}{2}}}
\;=\;
\frac{\sqrt{\pi}\,\Gamma(\frac{d}{2})}{2\,\Gamma(\frac{d+1}{2})}
\;,
\qquad
\mbox{Vol}(\SM^{d-1})
\;=\;
\frac{2\,\pi^{\frac{d}{2}}}{\Gamma(\frac{d}{2})}
\;,
$$
allows to complete the proof.
\hfill $\Box$

\vspace{.2cm}

\noindent {{\bf Acknowledgements:}  This work was partially supported by the DFG grant SCHU 1358/6-2. }


\end{document}